\documentclass[twocolumn,aps,pre,floatfix,superscriptaddress,longbibliography]{revtex4-1}
\pdfoutput=1 
\usepackage{amsmath,amssymb,eucal,graphicx,float,epstopdf,xparse}
\usepackage{epsfig,subfigure}
\usepackage[utf8]{inputenc}

\usepackage[colorlinks=true, urlcolor=blue, anchorcolor=blue, citecolor=blue,filecolor=blue,linkcolor=blue,menucolor=blue]{hyperref}

\def\bea{\begin{eqnarray}}
\def\eea{\end{eqnarray}}
\def\ba{\begin{eqnarray}}
\def\ea{\end{eqnarray}}
\def\beq{\begin{eqnarray}}
\def\eeq{\end{eqnarray}}
\def\be{\begin{equation}}
\def\ee{\end{equation}}
\def\eq{&=&}
\def\bm{\begin{math}}
\def\me{\end{math}}

\def\l{\left }
\def\r{\right }

\def\lb{\label}
\def\q{\quad}
\def\qq{\qquad}
\def\lb{\label}
\def\q{\quad}
\def\qq{\qquad}
\newcommand \nn {\nonumber}
\newcommand \bei {\begin{itemize}}
\newcommand \eei  {\end{itemize}}

\newcommand \nt   {\nonumber \\ }

\begin{document}

\title{Steady oscillations in aggregation-fragmentation processes}
\author{N. V. Brilliantov}
\affiliation{Department of Mathematics, University of Leicester, Leicester LE1 7RH, United Kingdom }
\affiliation{Skolkovo Institute of Science and Technology, Moscow, Russia}
\author{W. Otieno}
\affiliation{Department of Mathematics, University of Leicester, Leicester LE1 7RH, United Kingdom }
\author{S. A. Matveev}
\affiliation{Skolkovo Institute of Science and Technology, Moscow, Russia}
\author{A. P. Smirnov}
\affiliation{Faculty of Computational Mathematics and Cybernetics, Lomonosov MSU, Moscow Russia }
\affiliation{Institute of Numerical Mathematics RAS, Moscow, Russia }
\author{E. E. Tyrtyshnikov}
\affiliation{Faculty of Computational Mathematics and Cybernetics, Lomonosov MSU, Moscow Russia }
\affiliation{Institute of Numerical Mathematics RAS, Moscow, Russia}
\author{P. L. Krapivsky}
\affiliation{Department of Physics, Boston University, Boston, Massachusetts 02215, USA}

\begin{abstract}
We report surprising steady oscillations in aggregation-fragmentation processes. Oscillating solutions are observed for
the class of aggregation kernels $K_{i,j} = i^{\nu}j^{\mu} + j^{\nu}i^{\mu}$ homogeneous in masses $i$ and $j$ of
merging clusters and fragmentation kernels, $F_{ij}=\lambda K_{ij}$, with parameter $\lambda$ quantifying the intensity
of the disruptive impacts. We assume a complete decomposition (shattering) of colliding partners into monomers. 
We show that an assumption of a steady-state distribution of cluster sizes, compatible with
governing equations, yields a power-law with an exponential cutoff. This prediction agrees with simulations
results when $ \theta \equiv \nu-\mu <1$. For $ \theta=\nu-\mu >1$, however, the densities exhibit an oscillatory
behavior.  While these oscillations decay for not very small $\lambda$, they become steady if $\theta$ is close to two
and $\lambda$ is very small. Simulation results lead to a conjecture that for $ \theta <1$ the system has a stable
fixed point, corresponding to the steady-state density distribution, while for any $\theta >1 $ there exists a critical
value $\lambda_c(\theta )$, such that for $\lambda < \lambda_c(\theta)$, the system has an attracting limit cycle. This
is rather striking for a closed system of Smoluchowski-like equations, lacking any sinks and sources of mass.
\end{abstract}

\maketitle

\section{Introduction}

Numerous phenomena in nature involve dual processes of aggregation and fragmentation \cite{Krapivsky,Leyvraz}. These processes take place on vastly different length and time scales. A reversible polymerization in solutions and coagulation of colloidal particles are the classical examples of such processes occurring on the molecular scales; another peculiar example is aggregation of prions causing the Alzheimer-like diseases \cite{prions}. On the larger scales---in atmospheric processes, small airborne particles coalesce into smog droplets \cite{Srivastava1982}.  Aggregation is also common in systems of living organisms, from colonies of viruses \cite{Grant} to schools of fish \cite{Niwa}.  Aggregation and fragmentation processes occur in networks of different nature, including economic networks \cite{Takayasu} and internet communities \cite{Krapivsky,Dorogov}; here forums of users nucleate, merge and split. In turbulent cascades in a fluid flow \cite{Turbulence}, vortices may merge forming larger ones or decomposing into smaller vortices. The distribution of particles size in planetary rings is also determined by a steady balance achieved between two opposite processes, viz. aggregation and breakage of the particles in the rings \cite{PNAS,stadnichuk2015smoluchowski,Cuzzi,Brill2009,Esposito}.

\subsection{Aggregation}

\begin{figure}[h]
\begin{center}
\includegraphics[scale=0.11]{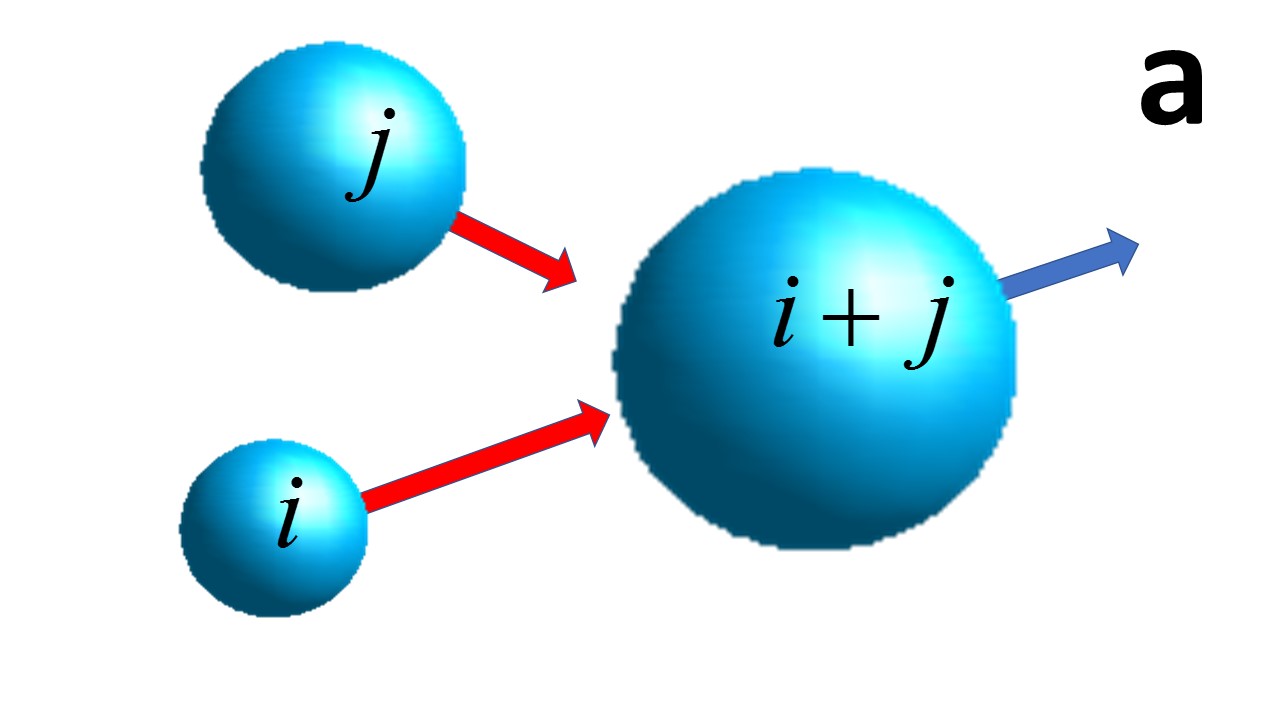} \qq
\includegraphics[scale=0.11]{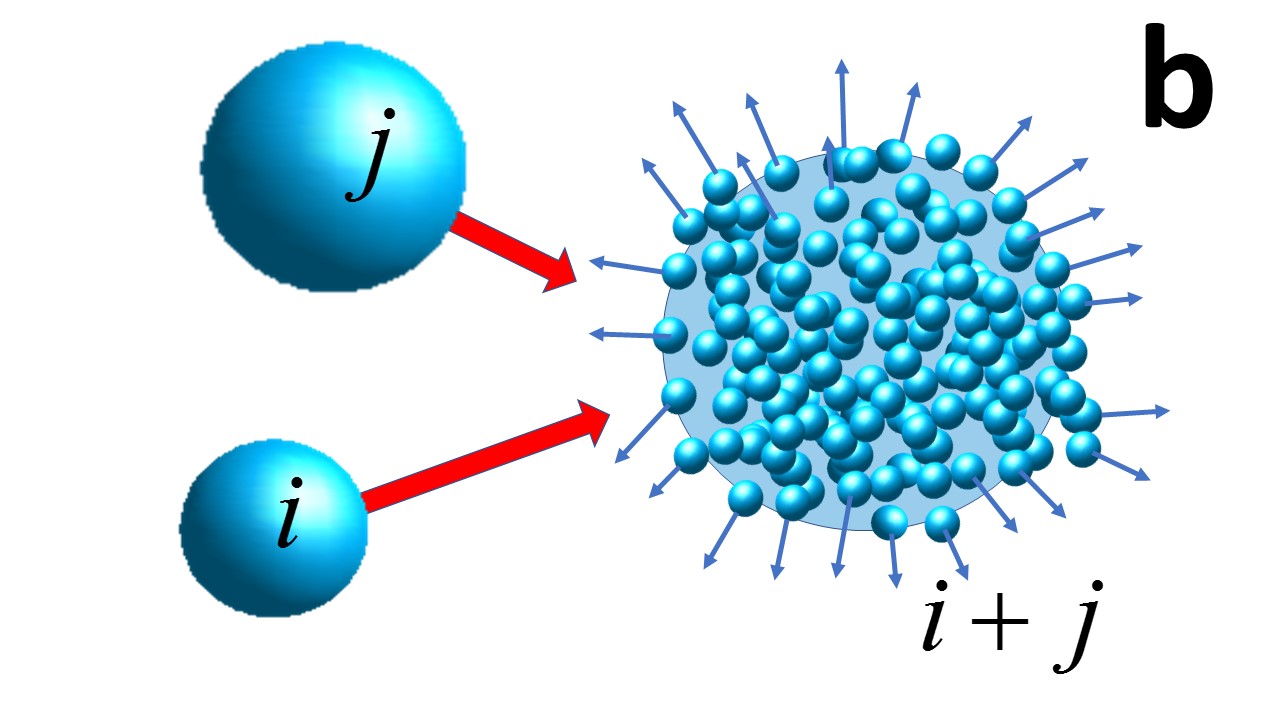}
\end{center}
\caption{(a) A merging event. (b) A collision of clusters of size $i$ and $j$ leading to decomposition
into $i+j$ monomers. }
\label{fig:Agg_Frag}
\end{figure}

The aggregation takes place when two clusters, comprised respectively of $i$ and $j$ monomers, merge upon collision thereby creating a cluster of $i+j$ monomers (see Fig.~\ref{fig:Agg_Frag}); symbolically this process
may be written as
$$
[i] +[j] \xrightarrow{K_{ij}} [i+j]
$$
where $K_{ij}$ is the merging rate. Let $n_k$ be the concentration of clusters of size $k$, i.e., clusters composed of 
$k$ monomers. The rate of change of $n_k$ is determined by Smoluchowski equations \cite{Krapivsky,Leyvraz}

\be
 \lb{eq:Smol1} \frac{dn_k}{dt} = \frac{1}{2}\sum_{i+j=k}K_{i,j}n_i n_j - n_k
\sum_{i=1}^{\infty}K_{i,k}n_i 
\ee
The first term in the right-hand side accounts for the formation rate of $k$-mers from clusters of size $i$ and
$j$, the second term describes the loss of $k$-mers due to aggregation of these clusters with
all other clusters; the factor $1/2$ in the first term prevents from double counting of the same process ($i+j
\to k$ and $ j+i \to k$).

\subsection{Aggregation with fragmentation}

Generally aggregates can suffer both spontaneous and collision fragmentation \cite{Krapivsky,Leyvraz,prions,Brill2009,PNAS,stadnichuk2015smoluchowski}. In the former case a cluster breaks into smaller pieces without interactions with other aggregates \cite{Krapivsky,Leyvraz,prions}, in the
latter one the fragmentation is caused by an energetic impact between two clusters \cite{Brill2009,PNAS,
stadnichuk2015smoluchowski}. Different collision fragmentation models have been studied~\cite{Brill2009,PNAS,
stadnichuk2015smoluchowski}; here we will consider a simple one of complete shattering of two colliding partners into
monomers. Symbolically this process (see Fig.~\ref{fig:Agg_Frag}) may be written as
$$
[i] +[j] \xrightarrow{F_{ij}} \underbrace{[1]+[1]+\ldots [1]}_{i+j}
$$
where $F_{ij}$ quantifies the shattering rate. Models with shattering exhibit interesting behaviors including dynamical phase transitions \cite{KOB2017}. 
It has been shown \cite{PNAS} that more general fragmentation models with large number of fragments yield qualitatively similar size distribution provided the small-size debris strongly dominates over the large size ones \cite{PNAS}.  As in Ref.~\cite{PNAS}, we assume that the fragmentation and aggregation kernels are proportional,
\be 
\lb{eq:CijAij}  F_{ij} = \lambda K_{ij},  
\ee
as has been justified for processes in planetary rings \cite{PNAS}. The parameter $\lambda$ in
Eq.~\eqref{eq:CijAij} characterizes the relative frequency of aggregative and shattering impacts.

Adding the fragmentation kinetics with kernel \eqref{eq:CijAij} into kinetic equations \eqref{eq:Smol1} we arrive at a separate equation
\beq 
\label{eq:n1} \frac{dn_1}{dt} \!\!\eq \!\! -n_1 \sum\limits_{i=1}^{\infty} K_{1,i} n_i + \frac{\lambda}{2}
\sum\limits_{i = 2}^{\infty} \sum\limits_{j=2}^{\infty} (i+j) K_{i,j} n_i~ n_j \nt &+& \lambda n_1
\sum\limits_{j=2}^{\infty} j K_{1,j} n_j
\eeq
for the concentration of monomers and a set of generic equations
\beq
\label{eq:nk} \frac{dn_k}{dt} \!\!\eq \!\!\frac{1}{2} \sum\limits_{i=1}^{k-1} K_{i,k-i} n_i n_{k-i} - (1 + \lambda) n_k
\sum\limits_{i=1}^{\infty} K_{k,i} n_i 
\eeq
for $k\geq 2$. The second term on the right-hand side of Eq.~(\ref{eq:n1}) accounts for the gain of monomers occurring in shattering collisions between clusters, the third term describes the gain of monomers in the shattering impacts between monomers and clusters. Equation \eqref{eq:nk} differs from \eqref{eq:Smol1} by an extra loss term (proportional to $\lambda$) accounting for shattering.

Equations \eqref{eq:n1}--\eqref{eq:nk} describe spatially-homogeneous systems. The kernels $K_{i,j}$ may be obtained from the microscopic analysis of the aggregation and fragmentation processes (see e.g. \cite{PNAS, Brill2009, BFP2018}). In applications,  $K_{i,j}$ are usually homogeneous functions of the masses $i$ and $j$ of merging clusters. 
Here we will investigate the kernels 
\be
\lb{eq:Cijgen} 
K_{i,j} = i^{\nu}j^{\mu} + i^{\mu}j^{\nu}
\ee
which are rather popular  \cite{Krapivsky,Leyvraz} and have been used for similar aggregating-shattering systems in Ref.~\cite{connaughton2016universality} where a source of monomers and sink of large aggregates was present. A stationary distribution satisfying Eqs.~\eqref{eq:n1}--\eqref{eq:nk} with the kernel \eqref{eq:Cijgen} has been also addressed in \cite{colmPRE2018}. 

In the following, we shall often use the sum and the difference of the exponents $\mu $ and $\nu$
\be \lb{eq:homnonl} 
\beta = \nu +\mu, \qq \theta = \nu -\mu .   
\ee
(Without loss of generality, we choose $\nu \geq \mu$.) The exponent $\beta$ is the well-known homogeneity exponent  \cite{Krapivsky,Leyvraz}. The exponent $\theta$ plays an important role in the following; it has been called a non-locality exponent in \cite{connaughton2016universality,colmPRE2018}. 

We always limit ourselves to the non-gelling case $\beta <1$.  The restrictions $\nu\leq 1$ and $\mu\leq 1$ are needed to avoid instantaneous gelation (see e.g. \cite{Ernst1983,van87,bk,Laurencot1999,Malyshkin2001}). The exponent $\theta$ can exceed 1, and never-ending oscillations are actually observed in a `non-local' regime $\theta>1$. 

In the special case of $\nu=-\mu=a$, the kernel reads
\be 
\lb{eq:Cija}
K_{i,j} = i^aj^{-a} + j^ai^{-a}
\ee
with $0\leq a \leq 1$. This kernel is known as a generalized Brownian kernel \cite{ColmPaulJCP2012}. In what follows we will analyze both \eqref{eq:Cijgen} and \eqref{eq:Cija}, often starting with the latter which is more tractable.  The restriction $a\leq 1$ is needed to avoid instantaneous gelation (aggregation equations with $a>1$ are ill-defined \cite{Ernst1983,van87,bk,Laurencot1999,Malyshkin2001,Colm2011}). The solutions of the
aggregation-fragmentation equations should also satisfy the natural physical requirement $n_k(t) \geq 0$,
and mass conservation:
\be 
\lb{eq:M1} 
M = \sum_{k=1}^{\infty} k n_k(t) \equiv {\rm const}. 
\ee
Analytical time-dependent solutions to Eqs.~\eqref{eq:n1}--\eqref{eq:nk}, have been obtained only for the simplest case of a constant kernel \cite{PNAS}. The steady-state solutions have been found for several other models, such as irreversible aggregation model with a monomer source \cite{hayakawa1987irreversible}, an
aggregation-fragmentation model with kernels $K_{i,j}= (ij)^{\mu}$ and $F_{i,j}= \lambda K_{i,j}$~\cite{PNAS}, and for an open aggregation-fragmentation system with a source of monomers and sink of large clusters~\cite{connaughton2016universality} for the kernels of the form \eqref{eq:Cijgen} and for closed systems in
\cite{colmPRE2018}. An {\em open} aggregating system with the same coagulation kernel \eqref{eq:Cijgen}, driven by input of monomers along with the removal of large clusters has been studied in \cite{Colm}. Steady oscillations were numerically found in this system with a finite number of aggregate species \cite{Colm}. For a closed system comprised of monomers, dimers, trimers and exited monomers, stable oscillations have been also reported \cite{Gorban}. Similarly, steady chemical oscillations may occur in a dimerization model (see, e.g., \cite{dimer}).

In the present study we consider {\em closed}  systems undergoing aggregation and fragmentation processes, with the kinetic rates given by Eqs.~\eqref{eq:Cijgen} and \eqref{eq:CijAij}, that lack any source or sink of monomers and clusters. Naively, one expects that such closed systems with two opposite processes will relax to a steady-state where a balance between aggregation and shattering is established. This scenario is indeed realized for $\theta <1$, or
$a<1/2$ in the case of the kernels \eqref{eq:Cija}. Unexpectedly,  for $\theta \to 2 $ (or $a \to 1$) and small values of $\lambda$ we observe never-ending oscillations of the concentrations. This effect has been found numerically for the one-parameter family of kernels \eqref{eq:Cija} and reported in our recent study \cite{Oscil1}. Here we present a more detailed analysis of the aggregating and shattering systems, both numerical and theoretical, and we investigate a more general two-parameter family of kernels \eqref{eq:Cijgen}. We also provide a qualitative theory of the stable oscillations which sheds some light on the
mechanism of this surprising phenomenon.

In what follows, we will concentrate on systems with time-independent coefficients $K_{ij}$ and $\lambda$, and conserved total mass (total number of elementary units). This is a generic model that describes systems of very different nature. The elements comprising a system range from grains or molecules to living organisms or economic agents. The interaction forces, that determine the kinetic rates, may be of very different nature as well. These may range from true molecular or mechanical forces to fictitious ``social forces" \cite{Helbing} based on informational exchange. Therefore, strictly speaking, the term ``closed" here literally means a lack of sinks and sources of system elements. At the same time, the exchange of energy, chemicals, nutrients, and information is implied.This is needed to sustain elements of a system and keep the rate coefficients steady. On the level of social agents or a living organism this implies an interaction with the surrounding social or natural environment. On the level of molecular or macroscopic particles the interaction with a thermostat, or the presence of some other source of energy, is assumed. For instance, in aggregation-fragmentation processes in polymer or colloidal solutions, there exists energy exchange with the solvent. This maintains constant temperature, although energy is released in aggregation processes and consumed in fragmentation processes. Similarly the surrounding molecular gas plays a role of thermostat in atmospheric processes \cite{Friedlander} and for dust clouds \cite{Ossenkopf,BrilliantovSpahn2006}. Another important mechanism of energy supply is viscous heating, which arises in planetary rings \cite{Esposito}. In this case the orbital motion of rings' particles yields a sheared flow of viscous granular fluid, which generates heat \cite{Esposito}. The energy supply keeps the kinetic energy of aggregates steady, and the rate coefficients constant.

Systems with true molecular or mechanical forces between elements is an important subclass of systems with aggregation and fragmentation. As it follows from the discussion above, such systems, with constant rate coefficients, are not thermo-dynamically closed. (Note that the notion ``thermodynamics" is meaningful only for these systems.) Hence an interesting question arises---whether persistent concentration oscillation sexist in thermodynamically closed systems? We perform a microscopic analysis, which resulted in a positive answer: Never-ending oscillations do emerge in thermodynamically closed systems, although the oscillation period permanently increases. 

The rest of the paper is organized as follows. In the next Sec.~\ref{sec:Numerics}, we present simulation results obtained with the use of fast solvers of Smoluchowski-type equations. In Sec.~\ref{sec:Theory} we discuss steady-state distributions using the methods outlined in
Ref.~\cite{Oscil1} and applied to Brownian kernels $\mu=-\nu =a$. We also present a qualitative theory explaining the
mechanism leading to never-ending  oscillations. In Sec.~\ref{sec:summary} we summarize our findings.

\section{Numerical results}
\label{sec:Numerics} 

Kinetic equations \eqref{eq:n1}--\eqref{eq:nk} form a set of infinitely many nonlinear coupled ordinary differential equations (ODE), which is a severe numerical challenge. For standard Smoluchowski equations, that is when fragmentation is absent, the average size of aggregates grows indefinitely imposing a time limit to model these processes. Fragmentation precludes the formation of very large clusters (in most cases and certainly in our case when fragmentation and aggregation kernels are proportional). This allows to model the aggregating-and-shattering systems with a finite number of equations $N_{\rm eq}$, which is dictated by the requested accuracy. In Ref.~\cite{Oscil1} we present estimates that relate the number of equations and the simulation accuracy; in practice we use such number of equations, that a further increase of $N_{\rm eq}$ does not impact the results for the concentrations $n_k(t)$ within the numerical precision.

The structure of the kinetic kernels \eqref{eq:Cija}  allows to apply highly efficient numerical methods, in particular, the fast and accurate method of time-integration of Smoluchowski-type equations \cite{matveev2015fast,Chaudhury2014, hackbusch2006efficient, hackbusch2007approximation,matveev_CPC2018}. The efficiency and accuracy of this approach in solving the aggregating-and-shattering equations has been demonstrated in Ref.~\cite{Oscil1}, where the numerical results have been compared with the available analytical
solutions~\cite{PNAS}.

\subsection{Steady state size distribution}

\begin{figure}[h!]
\begin{center}
\includegraphics[scale=0.6,page=4]{FIG2.pdf}
\includegraphics[scale=0.6,page=3]{FIG2.pdf}
\includegraphics[scale=0.6,page=2]{FIG2.pdf}
\includegraphics[scale=0.6,page=1]{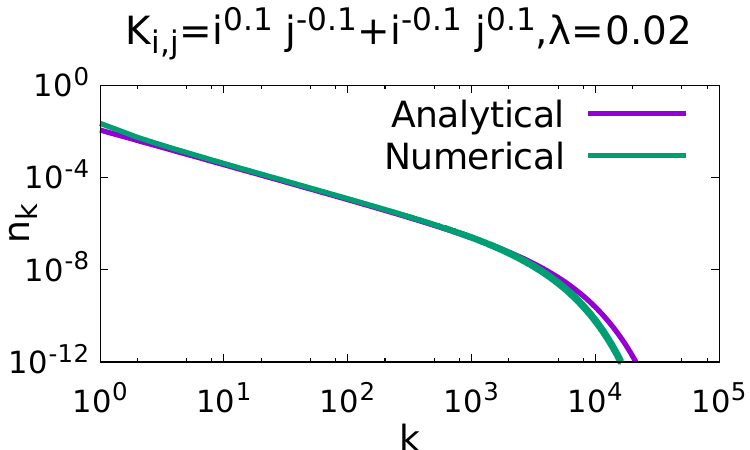}
\caption{Steady-state distributions obtained numerically by solving Eqs.~\eqref{eq:n1} and \eqref{eq:nk} for the Brownian kernel \eqref{eq:Cija}. Analytical results, Eq.~\eqref{eq:disfun2_a}, are also shown.} \label{pic:steady}
\end{center}
\end{figure}
Solving numerically Eqs.~\eqref{eq:n1}--\eqref{eq:nk} with kernel \eqref{eq:Cija} for
$a<1/2$,  we observe that the concentrations relax monotonically to a steady-state, see Fig.~\ref{pic:steady}.  In Fig.~\ref{pic:steady} we also compare the numerical results with the analytical solution for the steady-state distribution $n_k$, discussed below. The numerical and analytical solutions agree fairly well.

\begin{figure}[h!]
\begin{center}
\includegraphics[scale=0.6,page=2]{FIG3A_top.pdf}
\includegraphics[scale=0.6,page=4]{FIG3A_top.pdf}
\includegraphics[scale=0.6,page=2]{FIG3A_bot.pdf}
\includegraphics[scale=0.6,page=1]{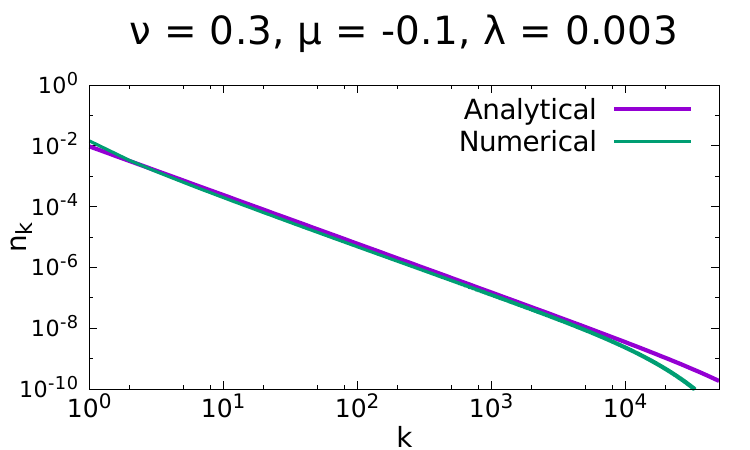}
\caption{Steady-state distributions obtained numerically by solving Eqs.~\eqref{eq:n1} and \eqref{eq:nk} for kernels \eqref{eq:Cijgen} with $\theta=\nu-\mu<1$. Analytical results, Eq.~\eqref{eq:disfun2}, are also shown. } 
\label{pic:steady_gen}
\end{center}
\end{figure}

Similar  behavior is observed for the general kernel \eqref{eq:Cijgen}. When $\theta=\mu-\nu <1$, the concentrations relax monotonically to a steady-state, and the final distribution agrees with the one predicted theoretically, see Fig.~\ref{pic:steady_gen}. The steady-state size distribution may be interpreted as a stable fixed point in the language of dynamical systems \cite{Strogatz}.

\subsection{Oscillating solutions}

\subsubsection{Brownian kernels ($\nu=-\mu=a$)}

\begin{figure}[h!]
\begin{center}
\includegraphics[scale=0.6]{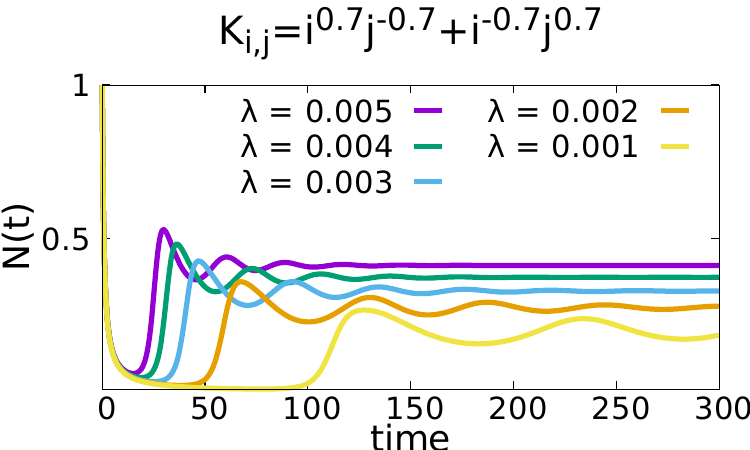} \qq
\includegraphics[scale=0.6]{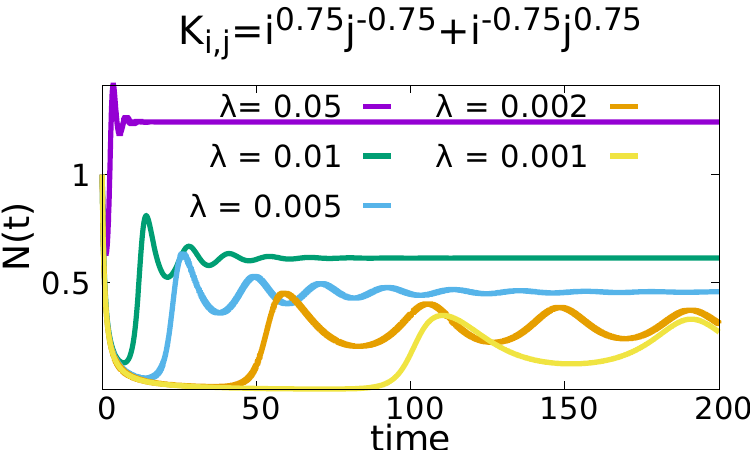}
\end{center}
\caption{Time dependence of the total density for $a=0.7$ (top) and $a=0.75$ (bottom) and different
$\lambda$. The system relaxes to a steady-state through damped oscillations which are more pronounced for larger
$a$ and smaller $\lambda$.  } \label{pic:Damping_osillations}
\end{figure}

For $a \geq 1/2$ a relaxation to a steady-state distribution occurs through oscillations, provided the parameter
$\lambda$, quantifying the shattering intensity, is relatively small. This is illustrated in
Fig.~\ref{pic:Damping_osillations}, where the time dependence of the total number of aggregates, $N(t)= \sum_{k\geq 1}
n_k(t)$, is shown; the figure also demonstrates that the oscillations are more pronounced and persist for longer time as
$a$ increases, while $\lambda$ decreases.

\begin{figure}[h!]
\begin{center}
\includegraphics[scale=0.6]{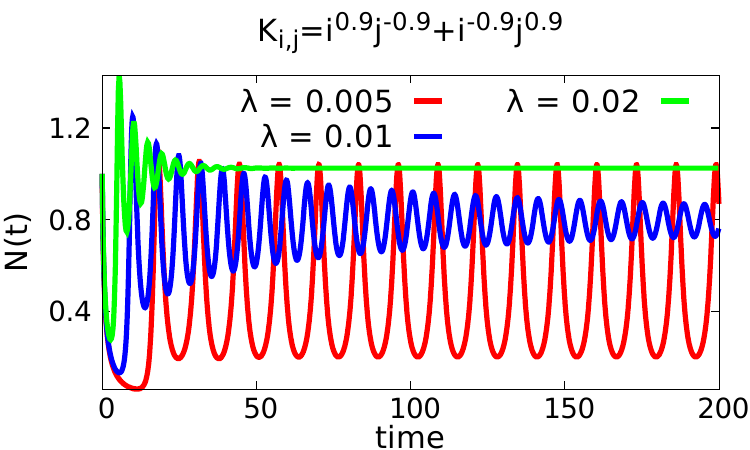}
\includegraphics[scale=0.6]{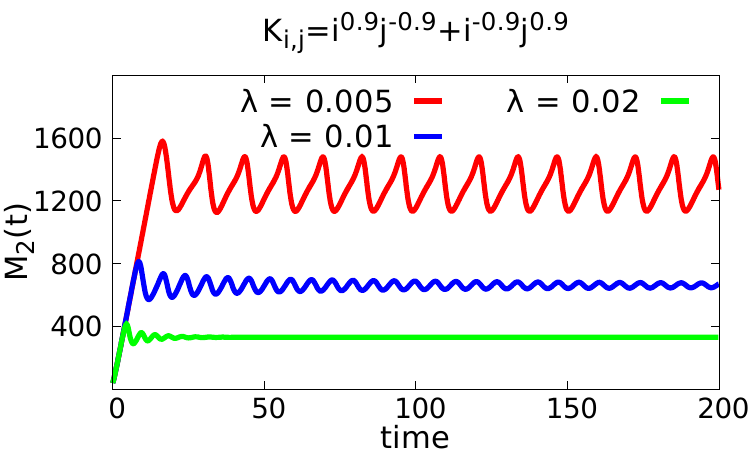}
\end{center}
\caption{Time dependence of the clusters density $N(t)$ (top) and the second moment $M_2(t)=\sum_{k \geq 1} k^2 n_k(t)$
(bottom), for the kernel \eqref{eq:Cija} with $a=0.9$  and different $\lambda$. Seemingly never-ending oscillations are observed for $\lambda=0.005$. }
\label{pic:Steady_osillations}
\end{figure}

We found the oscillations independently of initial conditions; here we use the mono-disperse initial
conditions, $n_k(0)=M\delta_{1,k}$ and step-wise initial conditions

\begin{align} \label{eq:Init_conditions}
n_k( t = 0) =\ \left\{\begin{matrix}
0.1 & k = 1, 2, \ldots 10\\
0   & k > 10,  \\
\end{matrix}\right.
\end{align}
with the same total mass $M = 5.5$. Unless explicitly stated, the reported results refer to the initial
conditions \eqref{eq:Init_conditions}. For $a \to 1$ and relatively small $\lambda$ we observe stable,
seemingly never-ending oscillations, see Fig.~\ref{pic:Steady_osillations}, where the temporal behaviors of the total density $N(t)$
and the second moment $M_2(t)= \sum_{k\geq 1} k^2 n_k(t)$ are depicted.

\begin{figure}[h!]
\begin{center}
\includegraphics[scale=0.6]{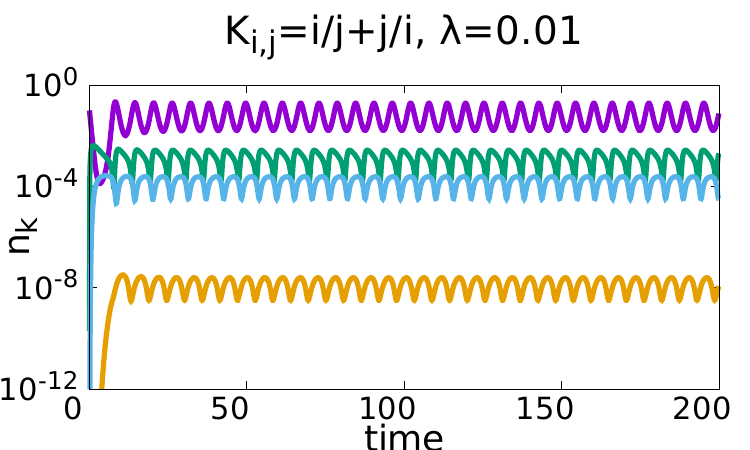} \qq
\includegraphics[scale=0.6]{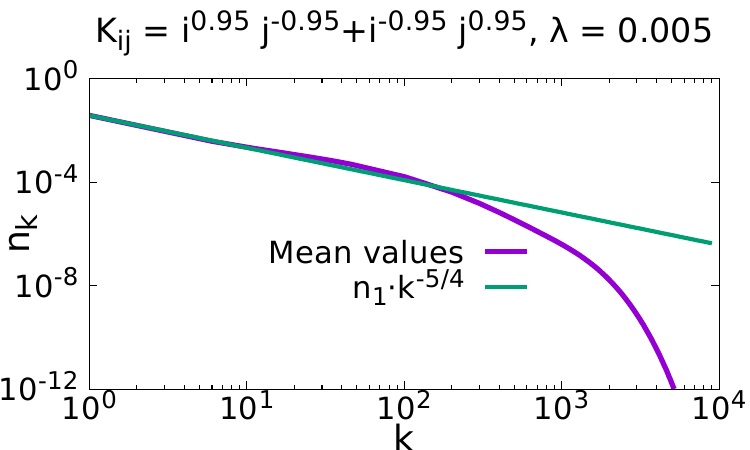}
\caption{Top: Stable oscillations. Bottom: The concentration distribution after
averaging over the oscillation period. The averaged concentrations follow a power-law distribution 
with exponent close to $5/4$ for not too large masses.} \label{pic:Oscila1}
\end{center}
\end{figure}

Making the time averaging of the densities over the oscillation period, one obtains the distribution of the averaged quantities $\left<n_k \right>_{\rm osc}$, which has a form of the power-law with a cutoff at $k \sim
k_0$, see Fig.~\ref{pic:Oscila1}:

\be \lb{eq:nkav} \left<n_k \right>_{\rm osc} \sim k^{-\alpha}, \qq \alpha \simeq 5/4, \qq k<k_0. \ee

\subsubsection{General kernels \eqref{eq:Cijgen}}

\begin{figure}[h!]
\begin{center}
\includegraphics[scale=0.6,page=5]{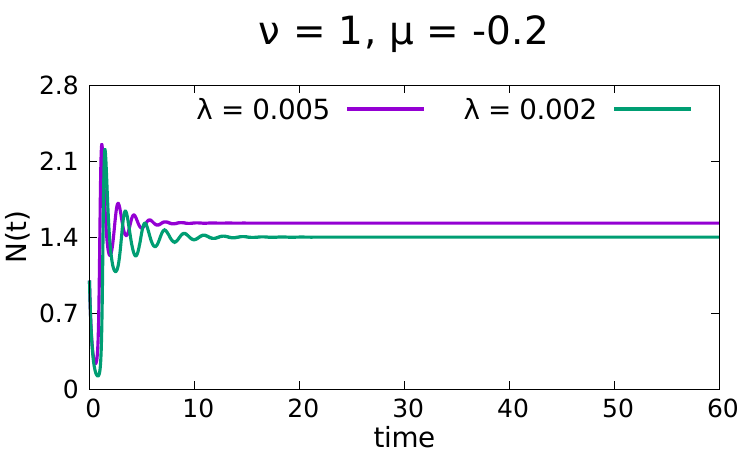}
\caption{The exponent $\theta=\nu-\mu=1.2>1$ is in a regime where never-ending oscillations are conjecturally possible for very small $\lambda$, but in the shown examples $\lambda$ is not small enough and the cluster size distribution relaxes to a
steady-state through the damped oscillations (which are more pronounced for smaller $\lambda$).}
\label{fig:Damping_osillations_gen}
\end{center}
\end{figure}

We observed oscillations for the general kernel \eqref{eq:Cijgen} when $\theta=\nu-\mu >1$ (which corresponds to $a>1/2$ of the Brownian kernel). However, if  $\theta$ is not close to $\theta=2$, the system relaxes to a steady distribution through the damped oscillations, even for rather small $\lambda$, see Fig.~\ref{fig:Damping_osillations_gen}.

\begin{figure}[h!]
\begin{center}
\includegraphics[scale=0.6,page=1]{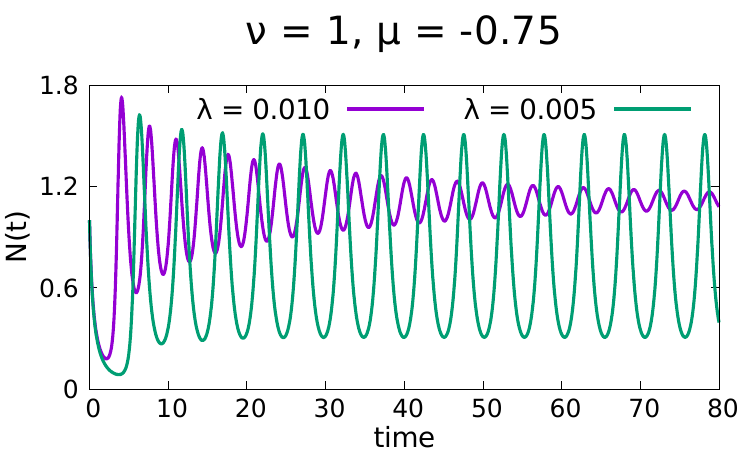}
\includegraphics[scale=0.6,page=2]{FIG6.pdf}
\includegraphics[scale=0.6,page=3]{FIG6.pdf}
\includegraphics[scale=0.6,page=4]{FIG6.pdf}
\caption{Oscillating behaviors when $\theta>1$. Never-ending oscillations emerge for rather small values of
$\lambda$; overall, the larger the exponent $\theta$, the larger the critical $\lambda_c(\theta)$ separating never-ending from damped oscillations.} \label{fig:Damping_osillations_gen_2}
\end{center}
\end{figure}

When $\theta\to 2$ (which corresponds to $a\to 1$), steady oscillations emerge for small
$\lambda$. Larger the exponent $\theta$, larger the shattering rate $\lambda$ where steady
oscillations emerge, see, Fig.~\ref{fig:Damping_osillations_gen_2}.

Our simulations imply the existence of a critical value $\lambda_c(\theta )$ such that for $\lambda < \lambda_c(\theta )$ in the long time limit the system approaches a limit cycle, viz. concentrations exhibit never-ending oscillations. This has
been checked for the Brownian kernel and for the general kernel \eqref{eq:Cijgen}. Although
for $a<0.9$ we have observed only damped oscillations, we believe that never-ending oscillations would emerge for
all $a>1/2$ and sufficiently small $\lambda$. This is seemingly true for the general case: The steady
oscillations would be observed for any $\theta>1$ if $\lambda$ is small enough. We cannot prove this
numerically due to unaccessible number of equations needed to simulate the systems with such small $\lambda$. For
instance, to simulate the system with $a=0.9$ and $\lambda =0.005$ depicted in Fig.~\ref{pic:Steady_osillations},  more
than 250 000 equations have been used. Our estimates (discussed in Ref.~\cite{Oscil1}) indicate that the number of
equations $N_{\rm eq}$, needed to guarantee a requested accuracy rapidly grows with the decreasing $\lambda$. To
simulate a system with $\lambda < \lambda_c$ for $a<0.9$ one needs more than a millions equations which is too
large for practical implementation. Nevertheless, based on our results, we
formulate the following 

\bigskip
\noindent{\bf Conjecture}. (i) When $\theta =\mu-\nu <1$, the system has a single stable fixed point for all values of $\lambda$; the steady state distribution of cluster sizes $n_k$ corresponds to this stable point. (ii) When $\theta >1$, there exists a critical
$\lambda_c(\theta)$, such that for $\lambda \geq \lambda_c$ the system possesses a stable fixed point with the
according distribution $n_k$. This may be a stable focus for some values of $\lambda$ manifesting in damped
oscillations. (iii) When $\theta >1$ and $\lambda < \lambda_c(\theta)$, the system possesses a
stable limit cycle.

As it follows from our numerical results, the critical shattering $\lambda_c$ strongly depends on the exponent $\theta$; its dependence on the other exponent $\beta=\mu+\nu$ seems to be weak (if any), but is still to be studied. 

\section{Theoretical analysis}
\label{sec:Theory}

To explain theoretically the observed behavior of the aggregation-and-shattering systems we analyze
separately the systems that attain a steady-state distribution and those that demonstrate never-ending
oscillations. For the former case we apply the asymptotic analysis, while in the
latter situation we analyze oscillations qualitatively.

\subsection{Asymptotic analysis of a steady-state cluster size distribution}

In Ref.~\cite{Oscil1} we gave a condensed account of the derivation of the steady-state
distribution; here we present  a more detailed derivation.

When the system reaches a steady-state, differential equations \eqref{eq:n1}--\eqref{eq:nk} become algebraic equations

\begin{eqnarray}
\label{eq:Model_stead}
&&n_1 \sum\limits_{i=1}^{\infty} K_{1,i} n_i - \lambda \sum_{i = 2}^{\infty} \sum_{j=2}^{\infty} (i+j) K_{i,j} n_i~ n_j
\\
&&~~~~~~~~~~~~~~~~~~~~~~~~~~~- \lambda n_1 \sum_{j=2}^{\infty} j K_{1,j} n_j =0\,, \nn \\
&& \frac{1}{2} \sum\limits_{i=1}^{k-1} K_{i,k-i} n_i n_{k-i} \!-\!
 (1 + \lambda) n_k \sum\limits_{i=1}^{\infty} K_{k,i}
n_i \!=\!0,  \q k \geq 2. \nonumber
\end{eqnarray}
To analyze these equations we introduce the generating functions 
\begin{equation}
\lb{eq:genfun} {\cal C}_{\gamma} (z)= \sum_{k=1}^{\infty} k^{\gamma} n_k z^k 
\end{equation}
and the moments
\begin{equation*}
{ M}_{\gamma} =
\sum_{k=1}^{\infty} k^{\gamma} n_k. 
\end{equation*}
Multiplying \eqref{eq:Model_stead} by $z^k$ and summing over all $k\geq 1$ we arrive at
\beq 
\lb{eq:gengen} 
&&{\cal C}_{\mu} (z){\cal C}_{\nu} (z) +(1+\lambda)zn_1({ M}_{\mu} + { M}_{\nu} )\\&& \qq -
(1+\lambda)\left(M_{\mu} {\cal C}_{\nu} (z) + M_{\nu} {\cal C}_{\mu} (z) \right)=0.  \nn\eeq
Using ${\cal C}_{\gamma} (1) = {M}_{\gamma}$ and specializing \eqref{eq:gengen} to $z=1$ we obtain

\be \lb{eq:MaMa} { M}_{\mu} { M}_{\nu} = \frac{1+\lambda}{1+2 \lambda} n_1 ({ M}_{\mu} + { M}_{\nu} ). \ee
To analyze  $n_k$ for $k \gg 1$ we will use the above equations and exploit standard methods of asymptotic analysis 
to extract the behavior of the generation functions ${\cal C}_{\gamma} (z)$. We consider
separately kernels with $\theta<1$ and $\theta>1$.

\subsubsection{Kernels with $\theta<1$ ($a<1/2$). }

It is known \cite{PNAS} that for $\mu=\nu=0$, the tail of the steady-state distribution reads $n_k  \simeq \lambda \pi^{-1/2} k^{-3/2}e^{-\lambda^2 k}$. Let us assume that for $k \gg 1$ our steady-state
distribution has a similar form:

\be \lb{eq:st_st} n_k  \simeq C k^{-\tau} e^{-\omega k }  \qq {\rm for}   \qq k \gg 1 \ee
with yet unknown $\tau$, $\omega$ and $C$. Equation \eqref{eq:st_st} implies

\be \lb{eq:GF} 
{\cal C}_{\gamma} (z)  \simeq \sum_{k =1}^{\infty}  C k^{\gamma-\tau}  (z/z_0)^k = \sum_{k =1}^{\infty}  C
k^{\gamma-\tau} (z^{\prime} )^k ,  \ee
where $z_0=e^{\omega}$ and $z^{\prime}=z/z_0$. Obviously, ${\cal C}_{\gamma}
(z^{\prime})$ diverges for $z^{\prime}>1$ and converges for $z^{\prime}<1$ for all $\gamma$ and $\tau$. We assume, that
${\cal C}_{\gamma} (z^{\prime}=1)$ exists, that is, $\sum_{k \geq 1}k^{\gamma-\tau}$ converges.

The tail of $n_k$ is reflected in the behavior of ${\cal C}_{\gamma} (z^{\prime})$ when $z^{\prime} \to 1-0$. Suppose $\sum_{k \geq 1}k^{\gamma-\tau+1}$ diverges.  Still, $\sum_{k \geq 1}k^{\gamma-\tau+1} (z^{\prime} )^k$ converges for $z^{\prime}<1$. The closer $z^{\prime}$ is  to $1$, the larger the size of the clusters $k$, that make the main contribution to  ${\cal C}_{\gamma} (z^{\prime})$. Hence the dependence  of ${\cal C}_{\gamma} (z^{\prime})$ on $z^{\prime}$ for $z^{\prime} \to 1$ characterizes the dependence of $n_k$ on $k$ for $k\gg 1$. To quantify this relation we differentiate ${\cal
C}_{\gamma} (z)$ with respect to $z$:
\begin{eqnarray*}
\frac{d {\cal C}_{\gamma}}{dz} &\simeq& C\, z_0^{-1}  \sum_{k=1}^{\infty} k^{\gamma-\tau+1} (z^{\prime}
)^{k+1} \\ 
&\simeq& C\, z_0^{-1} \int_0^{\infty} dk\, k^{\gamma-\tau+1} e^{k\log z^{\prime}} \\
&\simeq& C\, z_0^{-1} \int_0^{\infty} dk\,
k^{\gamma-\tau+1}   e^{-k (1-z^{\prime})} \\
& = & C\, z_0^{-1}  \Gamma(\gamma-\tau+2) (1-z^{\prime})^{\tau-\gamma-2} 
\end{eqnarray*}
where $\Gamma(x)$ is the gamma-function and we use $z^{\prime} \to 1-0$. Integrating with respect to $z$ we
obtain
\be \lb{eq:Gen_f_exp1} {\cal C}_{\gamma} (z) = {\cal C}_{\gamma} (z_0) + C \Gamma(1+\gamma -\tau) (1-
z^{\prime})^{\tau-\gamma-1} . \ee
Substituting ${\cal C}_{\gamma} (z)$ with $\gamma=\nu$ and $\gamma=\mu$ into Eq.~\eqref{eq:gengen} we
obtain terms with different powers of $(1-z^{\prime})$. To satisfy this equation we equate to zero all
these terms separately. The zero-order terms yield
\beq 
\lb{eq:zero} {\cal C}_{\nu} (z_0){\cal C}_{\mu} (z_0) &-&(1+\lambda) \left( {M}_{\nu}{\cal C}_{\mu} (z_0) +
{M}_{\mu} {\cal C}_{\nu} (z_0) \right) \nn \\& +& (1+\lambda)z_0 n_1 ({M}_{\nu} + {M}_{\mu} )=0.  
\eeq
The terms of the order $(1-z^{\prime})^{\tau +\gamma -1}$ with $\gamma=\nu$ and $\gamma=\mu$ imply
\be 
\lb{eq:pma_1} 
{\cal C}_{\mu } (z_0) C \Gamma(1 +\nu -\tau) -(1+\lambda) {M}_{\mu }  C \Gamma(1 +\nu -\tau)=0
\ee
and
\be 
\lb{eq:pma_2} 
{\cal C}_{\nu } (z_0) C \Gamma(1 +\mu -\tau) -(1+\lambda) {M}_{\nu }  C \Gamma(1 +\mu -\tau)=0
\ee
Finally, the rest of the terms should satisfy

\beq \lb{eq:fineq} && C^2 \Gamma(1 + \nu -\tau) \Gamma(1 +\mu  -\tau) (1-z^{\prime})^{2\tau -\nu-\mu -2 } \nt &&
~~~~~~~-(1+\lambda)z_0 n_1 ({M}_{\nu} + {M}_{\mu} ) (1-z^{\prime}) =0  \eeq
from which $2\tau -\nu-\mu-2 = 1$, or 

\be \lb{eq:tau} \tau =\frac{3+\beta}{2}\ee
where $\beta=\nu+\mu$. Now we substitute 

\be \lb{eq:MaCa} {\cal C}_{\gamma } (z_0) =(1+\lambda) {M}_{\gamma },  \ee
which follows from \eqref{eq:pma_1} and  \eqref{eq:pma_2} into \eqref{eq:zero} to obtain

\be \lb{eq:MaMa1} {M}_{\nu} {M}_{\mu} = \frac{z_0 }{1+  \lambda} n_1 ({M}_{\nu} + {M}_{\mu} ). \ee
From Eqs.~\eqref{eq:MaMa1} and \eqref{eq:MaMa} we get 

\be \lb{eq:z0} z_0 =e^{\omega} = \frac{(1+\lambda)^2}{(1 +2 \lambda)}.\ee
We have $\omega \simeq \lambda^2 - 2 \lambda^3 + \ldots \simeq \lambda^2$ for small $\lambda$ leading to 
\be 
\lb{eq:disfun2} 
n_k \simeq \frac{C}{k^{(3+\beta)/2} } e^{-\lambda^2 k}   \qq {\rm for}   \qq k \gg 1.\ee
To estimate the constant $C$ we utilize the distribution \eqref{eq:disfun2} together with mass conservation to 
yield
\begin{eqnarray}
\lb{eq:norm}
M = \sum_{k=1}^{\infty} k n_k &\approx&  \int_1^{\infty} dk\, \frac{C}{ k^{(1+\beta)/2}} e^{-\lambda^2 k}\\
& \simeq &  C \lambda^{\beta-1} \Gamma \left( \frac{1-\beta}{2} \right) \nonumber
\end{eqnarray}
resulting in
\be 
C \simeq \frac{M \lambda^{1-\beta} }{\Gamma\left[(1-\beta)/2\right]}\sim \lambda^{1-\beta} M. 
\ee
In the non-gelling $\beta<1$ region, the major contribution to the integral in \eqref{eq:norm} comes from $k\gg 1$, so the usage of \eqref{eq:disfun2} is justified.  For the Brownian kernel $\nu=-\mu =a$ and $\beta=0$, so the amplitude is $C \simeq  \lambda M/\sqrt{\pi} $ and

\be 
\lb{eq:disfun2_a} 
n_k \simeq \frac{\lambda M}{\sqrt{\pi} k^{3/2} } e^{-\lambda^2 k}   \qq {\rm for} \qq k \gg 1.
\ee

\subsubsection{Kernels with $\theta >1$ ($a > 1/2$). }

Applying the same analysis  for $\theta \geq 1$ (or $a \geq 1/2$), one arrives at Eqs.~\eqref{eq:zero}--\eqref{eq:fineq}, which however do not lead to consistent results. Indeed, from \eqref{eq:fineq} it follows that $\tau =(3+\nu+\mu)/2$. Substituting $\tau =(3+\nu+\mu)/2$ into Eq.~\eqref{eq:GF} we conclude that the generation
functions ${\cal C}_{\nu} (z_0)$ converges only for  $\nu -\mu =\theta <1$ (recall that $\nu \geq \mu$). Hence
Eq.~\eqref{eq:pma_1} may not be satisfied to cancel the terms corresponding to the factor $(1-z^{\prime})^{\tau +
\nu-1}$. The failure of this asymptotic analysis seemingly manifests the change of the evolution regime, which has been
observed in the numerical simulations:~For $ \theta >1 $ the oscillations of concentrations emerge --  the systems
either relax to a steady-state through damped oscillations, or demonstrate never-ending oscillations.

\subsection{Qualitative analysis}

To understand the mechanism of the stable oscillations let us consider a special Brownian kernel with $\nu=-\mu=a=1$. The  monomer density satisfies 
\begin{equation}
\lb{eq:2} 
\frac{dn_{1} }{dt} =  \lambda(N+M_2M_{-1} ) -(1+ \lambda)n_1(1+ M_{-1})
\end{equation}
and the rate equation for the cluster density is
\beq
\lb{eq:2a}
\frac{dN}{dt} =  \lambda(N+M_2M_{-1} ) -(1+2 \lambda) M_{-1}.
 \eeq
(We set $M=1$). These equations are not closed as they involve the moments $M_{-1}$ and $M_2$.  One can write the rate equation for $M_2$, but it involves the third moment $M_3$. This continues ad infinitum leading to analytically unsolvable hierarchy.

Note that on the right-hand side of Eq.~\eqref{eq:2} one term is negative and the other one is positive; the first is
of the order of $n_1$ and the second is of the order $\lambda M_2$. Initially only small clusters present in the
system, so that $\lambda M_2$ is small for $\lambda \ll 1$ and $n_1$ decreases. Due to the conservation of mass the
decrease of $n_1$ implies the increase of other  concentrations. Hence, after some time a wider cluster size
distribution is established, such that $\lambda M_2$ increases. When it exceeds $n_1$, the right-hand side  of
Eq.~\eqref{eq:2} becomes positive and $n_1$ starts to grow. Due to the conservation of mass, the growth of $n_1$
implies the decay of other concentrations, which leads to the decrease of $M_2$ and eventually to the negative sign of
the right-hand side of Eq.~\eqref{eq:2}. Then the cycle repeats.

\begin{figure}[h!]
\begin{center}
\includegraphics[scale=0.51]{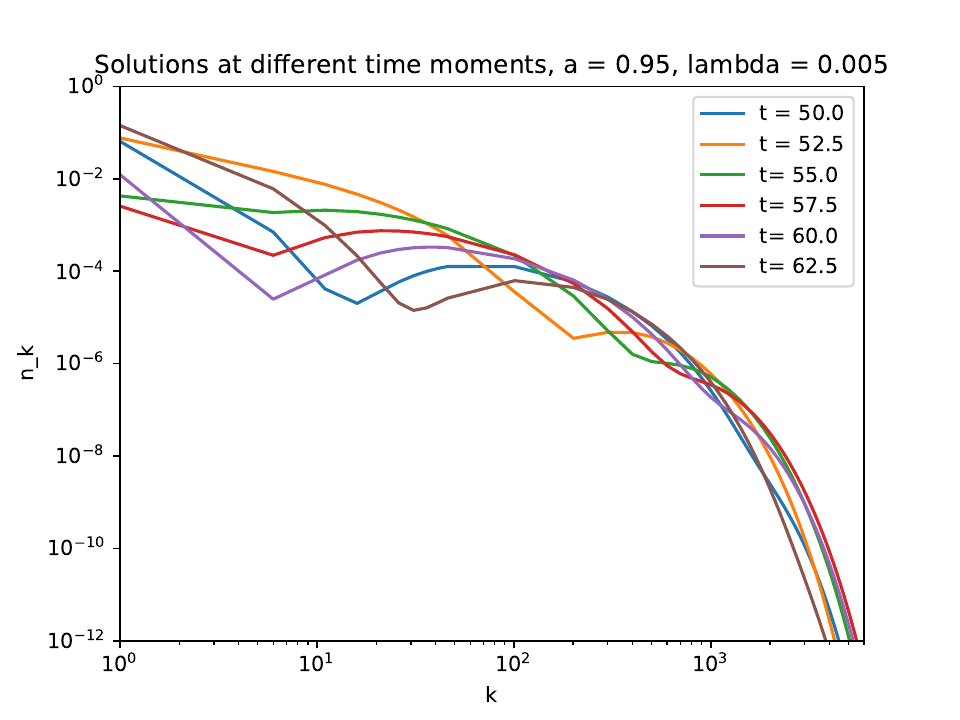}
\includegraphics[scale=0.58]{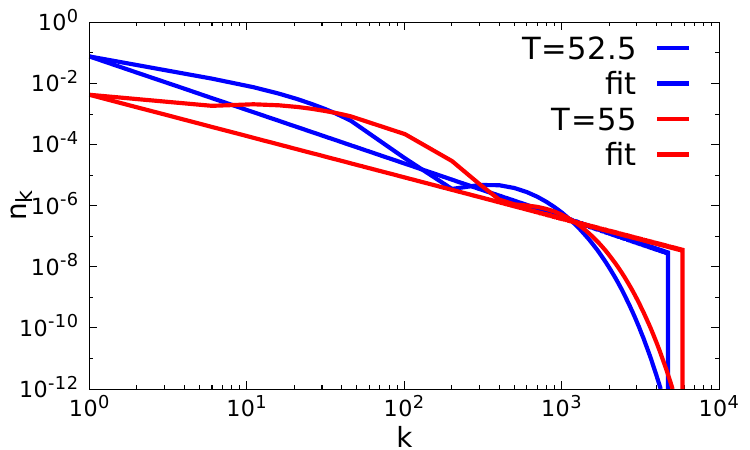}
\includegraphics[scale=0.58]{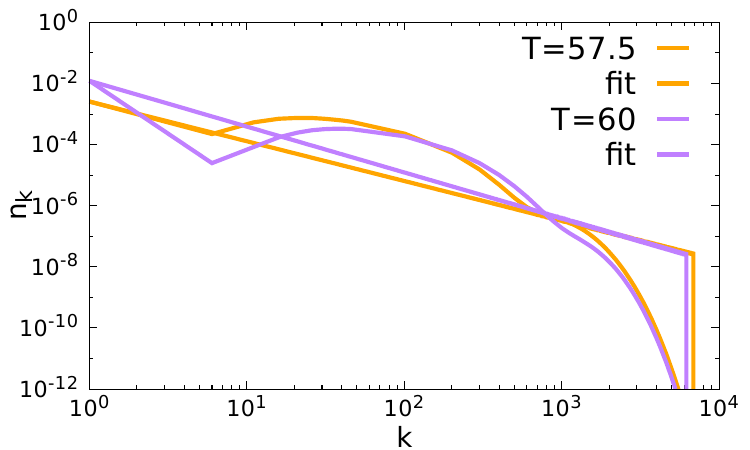}
\caption{Top panel: Time dependence of $n_k(t)$. The effective slope of the distribution and the effective cutoff size $k_{\rm max}$ periodically change with time. Middle and bottom panels: The real and coarse-grained cluster size distributions. The smooth cutoff of
the real distribution is approximated by the abrupt model cutoff. The simulation data are shown for the  kernel \eqref{eq:Cija} with $a=0.95$ and
$\lambda = 0.005$.} \label{pic:oscModel}
\end{center}
\end{figure}

Let us try to put the above narrative picture into somewhat more quantitative terms. Firstly, we notice that the
oscillations of  concentrations correspond to the periodically varying distribution of
cluster sizes, as it is illustrated in Fig.~\ref{pic:oscModel}. Roughly speaking, the size distribution
$n_k(t)$ behaves in such a way that the effective slope of this distribution $\alpha(t)$ and the effective
cutoff $k_{\rm max}(t)$ periodically change in time. Averaging over these oscillations we obtain the 
distribution $\left< n_k \right>_{\rm osc}$ depicted in Fig.~\ref{pic:Oscila1}.

\begin{figure}[h!]
\begin{center}
\includegraphics[scale=0.65]{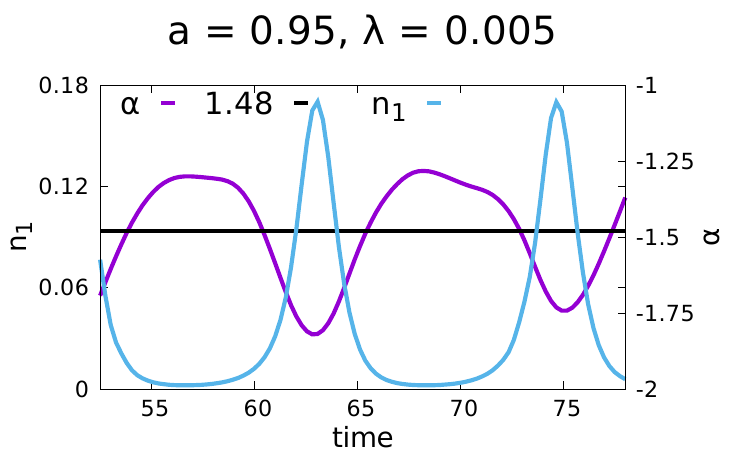} 
\caption{Periodic variation of the model parameters $\alpha(t)$ and $n_1(t)$.
Only two periods of oscillations are shown.  The simulation data are the same as for Fig.~\ref{pic:oscModel}.}
\label{pic:osc_n1_alpha}
\end{center}
\end{figure}

To understand the nature of the observed behavior of the system, we develop a qualitative theory. To this end
we approximate the real distribution $n_k(t)$ by a model distribution $n_k^{\rm
mod}(t)$ which reflects the most prominent features of the real distribution. Namely we assume that $n_k(t)$
may be characterized by a power-law distribution with a varying slope $\alpha(t)$ and a large-size cutoff
$k_{\rm max}(t)$; it should obey the mass conservation. Namely, we assume the following model distribution:
\begin{equation}
\label{eq:1} n_k^{\rm mod}(t)  
\begin{cases}
      \frac{n_1(t)}{ k^{\alpha(t)} } & \text{for $ k \leq k_{\rm max}(t) $} \\
        0                         & \text{ for $ k > k_{\rm max}(t).$}
\end{cases}
\end{equation}
The real distribution of the aggregate sizes $n_k(t)$ may be approximated by the model distribution
\eqref{eq:1} applying a coarse graining. For the qualitative analysis addressed here we exploited the most
simple approach. Namely, we use the value of $n_1(t)$, obtained in the simulations, and find the parameters
$\alpha(t) $ and $k_{\rm max}(t) $ from the conservation of mass. That is, we numerically find the pair
$(\alpha, \, \, \, k_{\rm max})$ with integer $k_{\rm max}$, that minimizes the difference, $|n_1\, H_{k_{\rm
max},\alpha}-1|$, where $H_{k_{\rm max},\alpha} =\sum_{i=1}^{k_{\rm max}} i^{-\alpha}$ are the generalized
harmonic numbers. The variation with time of the model distribution \eqref{eq:1}, which mimics the real distribution, is shown in Fig.~\ref{pic:oscModel}. In Fig.~\ref{pic:osc_n1_alpha} we show the periodic variation on the model parameters  $\alpha(t)$ and $n_1(t)$ and demonstrate that the variation of the slope is limited by the interval  $1 < \alpha(t)  <2$.

To perform a qualitative analysis we focus on the qualitative dependence of the moments $M_{-1}$, $N$, $M_2$ on $n_1$, $\alpha$, $k_{\rm max}$ and approximate the summation by integration:
\be 
\lb{eq:moments} 
M_i= \sum_{k=1}^{k_{\rm max}} k^i n_k \simeq  b_i \int_1^{k_{\rm max}} k^i n_k dk, 
\ee
where we introduce the coefficients $b_i$. These coefficients (assumed to be constant) are of the order of
one and account for the difference between integration and summation. Hence we obtain
\begin{equation}
\lb{eq:modmom} 
\begin{split}
&M_{-1}(t) \simeq  b_{-1}\frac{n_1(t)}{\alpha(t)},  \quad N(t)\equiv M_0(t) \simeq b_0\frac{n_1(t)}{\alpha(t)-1}\\
& M_2(t) \simeq  b_2\frac{n_1(t)}{3-\alpha(t)} \left[k_{\rm max}(t)\right]^{3-\alpha(t)}
\end{split}
\end{equation}
where we have used the condition $k_{\rm max} \gg 1$.  Similarly, the conservation of mass yields the relation
between $n_1(t)$, $\alpha (t)$, and $k_{\rm max}(t)$:

\be \lb{eq:n1kamx} M= \sum_{k=1}^{k_{\rm max}} k n_k \simeq  b_1 n_1 k_{\rm max}^{2-\alpha} (2-\alpha)^{-1}
=1. \ee

Using \eqref{eq:modmom}--\eqref{eq:n1kamx}, we recast \eqref{eq:2}--\eqref{eq:2a} into 

\begin{widetext}

\begin{subequations}
\begin{align}
\lb{eq:3} 
&\dot{n}_1 \eq \lambda n_1\left( \frac{b_0}{\alpha-1} + \frac{b_2b_{-1}n_1}{(3-\alpha) \alpha} \left(
\frac{2-\alpha}{b_1n_1} \right)^{\frac{3-\alpha}{2-\alpha}} \right) -
(1+\lambda)n_1\left(1+\frac{b_{-1}n_1}{\alpha}\right)\\
\lb{eq:3a} 
&\dot{\alpha} \eq \lambda (b_0+1-\alpha)\left( 1+ \frac{b_2 b_{-1}(\alpha-1)n_1}{b_0(3-\alpha) \alpha} \left(
\frac{2-\alpha}{b_1n_1} \right)^{\frac{3-\alpha}{2-\alpha}} \right) + \frac{(\alpha-1)^2(1+\lambda)}{\alpha} \left[
\frac{(1+2 \lambda) b_{-1}}{(1+\lambda)b_0} - \frac{\alpha +b_{-1}n_1}{\alpha-1} \right]
\end{align}
\end{subequations}

\end{widetext}

To show that never-ending oscillations are possible we perform the linear stability analysis of Eqs.~\eqref{eq:3}--\eqref{eq:3a}. We consider the coefficients $b_i$
($i=-1,\,0,\,1,\,2$) as known and of the order of unity. Further, we assume that there is  a fixed point, $n_1=n_{1}^{(0)}$ and $\alpha=\alpha_0$. At the fixed point $G_1(n_1^{(0)}, \alpha_0)=G_2(n_1^{(0)}, \alpha_0) =0$  where $G_1(n_1, \alpha)$ and for $G_w(n_1, \alpha)$ denote the right-hand sides of \eqref{eq:3} and \eqref{eq:3a}, respectively.  We also shortly write
\beq  
\lb{eq:4}
 g_{1n} \eq  \left. \frac{\partial G_1}{\partial n_1}  \right|_{n_{1}^{(0)}, \alpha_0}
\qq \qq g_{1\alpha }= \left. \frac{\partial G_1}{\partial \alpha }  \right|_{n_{1}^{(0)}, \alpha_0} \\
g_{2n} \eq  \left. \frac{\partial G_2}{\partial n_1}  \right|_{n_{1}^{(0)}, \alpha_0} \qq \qq g_{2\alpha }= \left.
\frac{\partial G_2}{\partial \alpha }  \right|_{n_{1}^{(0)}, \alpha_0}   \nn  \eeq
and deduce the linearized equations
\begin{equation}
\label{mat}
\frac{d}{dt}\,\left( \begin{array}{c}
\delta n_1 \\
\delta \alpha
\end{array} \right) =  \left( \begin{array}{cc}
g_{1n} &  g_{1\alpha}\\
 g_{2n} &  g_{2\alpha}
 \end{array} \right)
 \left( \begin{array}{c}
\delta n_1 \\
\delta \alpha
\end{array} \right) 
\end{equation}
for the deviations $\delta n_1 = n_1-n_{1}^{(0)}$ and $\delta \alpha = \alpha- \alpha_0$. The eigenvalues of the 
matrix in \eqref{mat} are
\be 
\lb{eq:eigval} 
\nu_{1,2} = \frac12 \left[g_{1n}+g_{2\alpha}\pm \sqrt{ (g_{1n} -
g_{2\alpha})^2+4g_{1\alpha}g_{2n} }\right]. 
\ee
Oscillations may occur if the above eigenvalues possess an imaginary part. This condition, ${\rm Im}(\nu_{1/2}) \neq 0$, require the negatives determinant in
\be 
\lb{eq:Dneg} 
D=(g_{1n} - g_{2\alpha})^2+4g_{1\alpha}g_{2n} <0. 
\ee
If the real part of the eigenvalues is negative, that is $(g_{1n}+g_{2\alpha})/2<0$, the fixed point is stable; in this case the cluster distribution relaxes to the stead-state $n_k$. In the opposite case of positive real part, $(g_{1n}+g_{2\alpha})/2>0$, the fixed point is linearly unstable and the oscillations grow and eventually stabilized by non-linear terms. 

\begin{widetext}

\begin{figure}[h!]
\begin{center}
\includegraphics[scale=0.55]{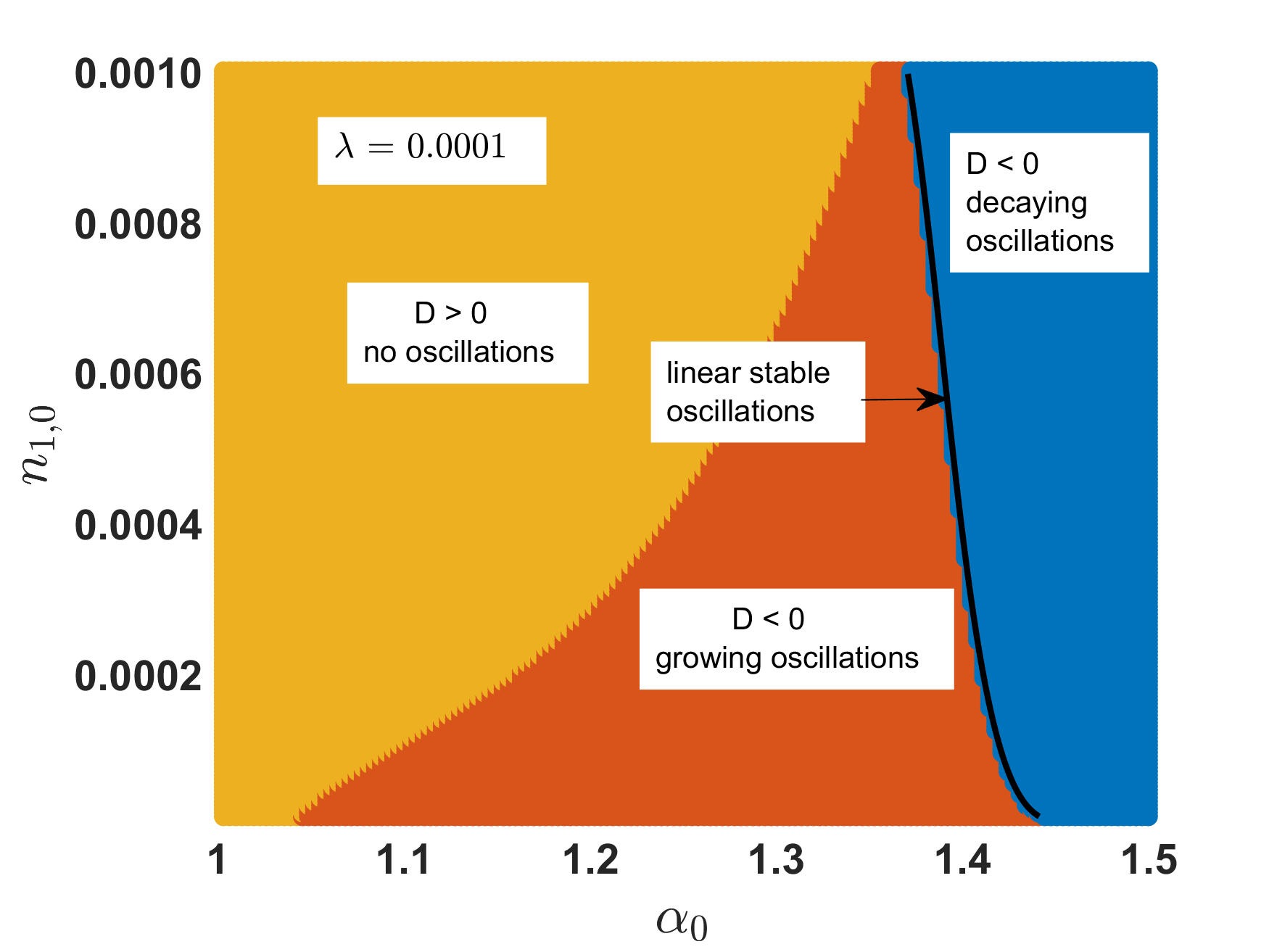} \q
\includegraphics[scale=0.55]{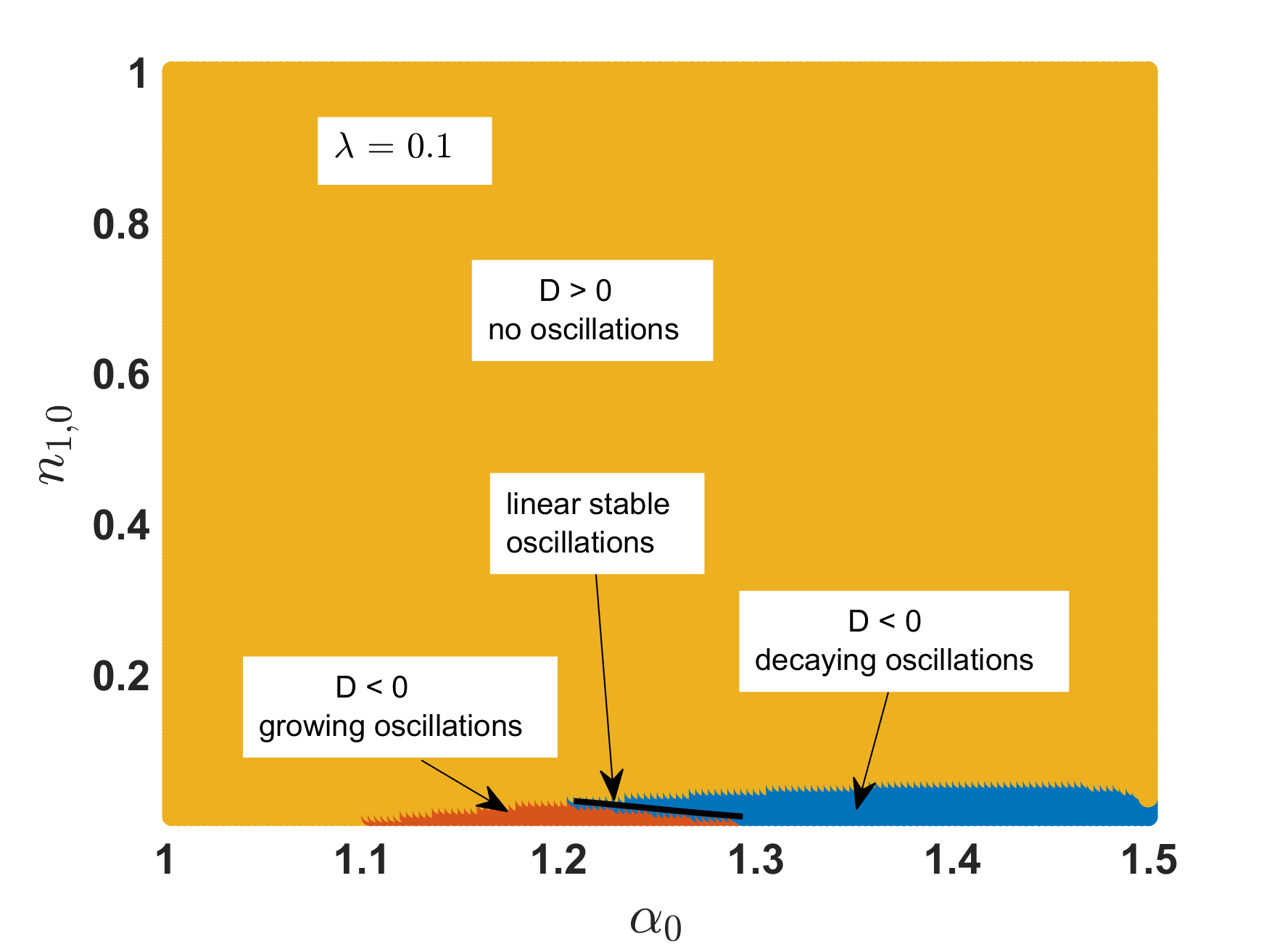} \\
\includegraphics[scale=0.55]{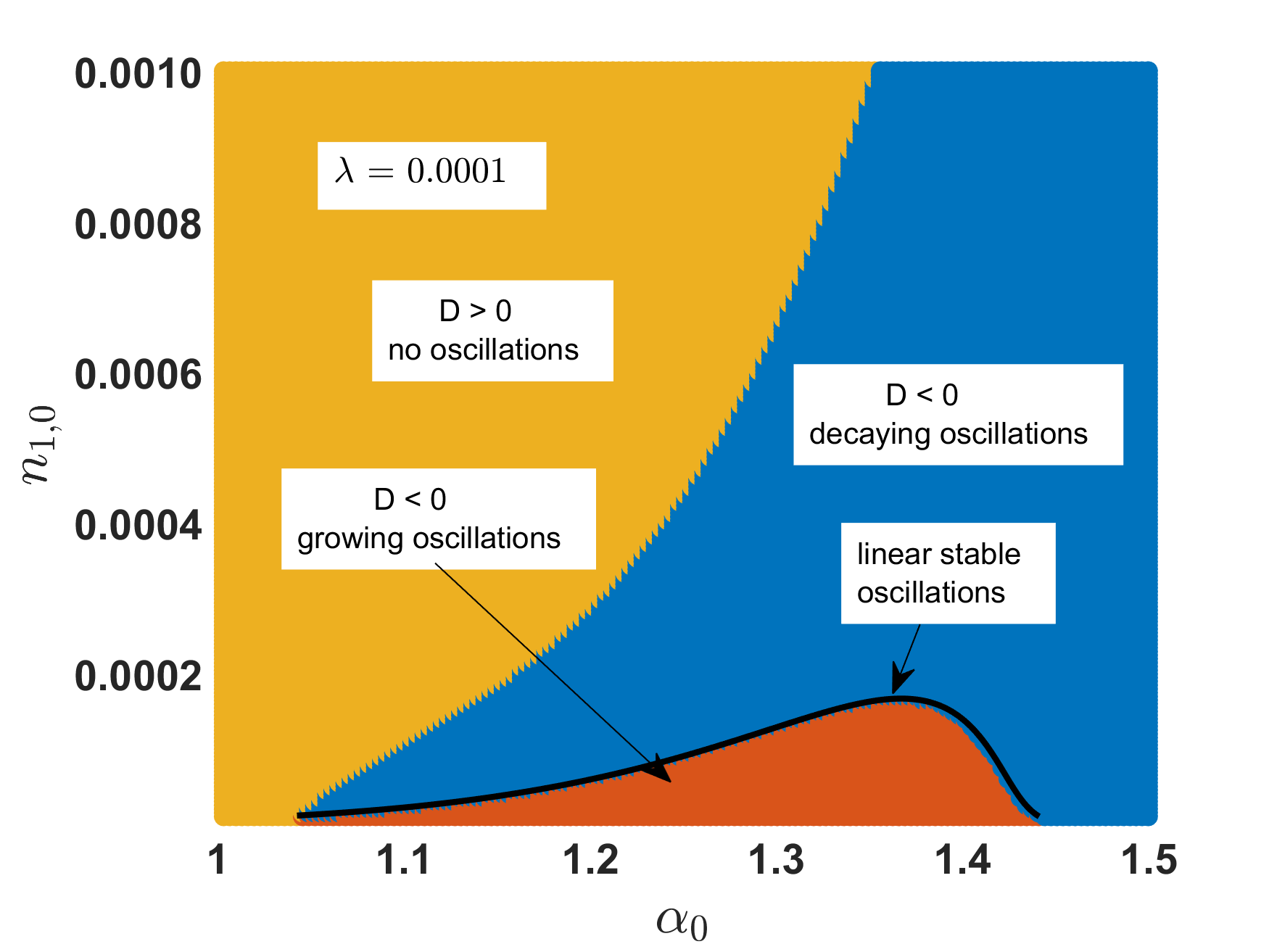} \q
\includegraphics[scale=0.55]{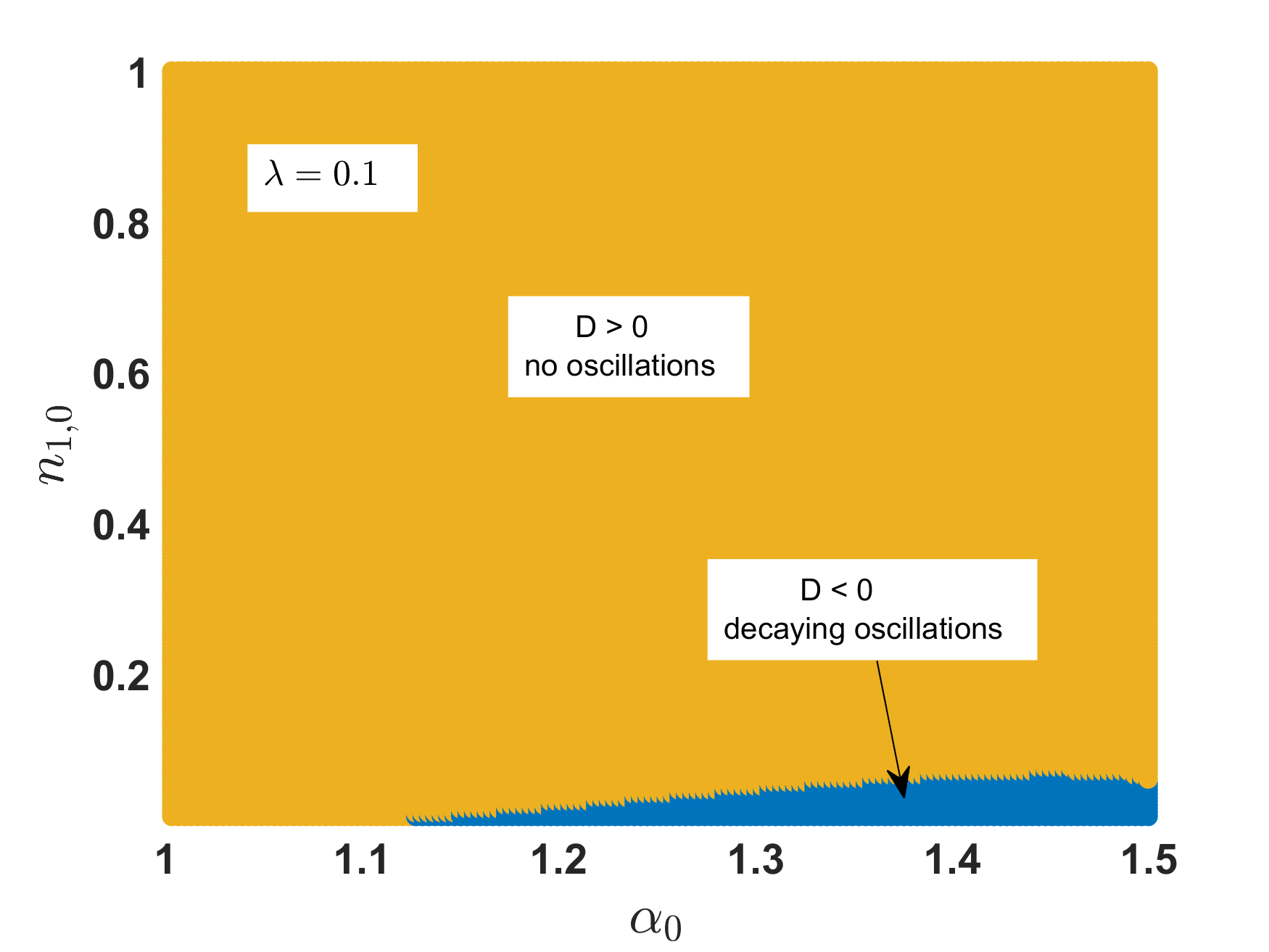}
\caption{Kinetic regimes for different location of the fixed point,  ($\alpha_0$, $n_1^{(0)}$), as it follows from
Eqs.~\eqref{eq:eigval}, \eqref{eq:Dneg} for $\lambda =0.0001$ (left panels) and $\lambda=0.1$ (right panels) and the
coefficients $b_{-1} = 2$, $ b_0 = 0.5$, $b_1 = 1$, $b_2 = 1$ (upper row) and $b_{-1} = 0.25$, $ b_0 = 0.5$, $b_1 = 1$,
$b_2 = 1$ (bottom row). For the negative determinant ($D<0$) the oscillations emerge. The stability of the oscillations
is determined by the sign of $g_{1n}+g_{2\alpha}$. Note that for large $\lambda=0.1$ (right panels) stable oscillation,
both linearly stable and linearly unstable (with the non-linear stabilization), may arise in a very small area of the
parametric space, while for small $\lambda=0.0001$ (left panels) the according domain is rather large.}
\label{fig:Zones}
\end{center}
\end{figure}

\end{widetext}

The coefficients $b_i$ are unknown, so we cannot locate the fixed point $(n_{1}^{(0)}, \alpha_0)$. Numerically we observe that $b_i=O(1)$;  the location of the fixed point corresponds to $n_{1}^{(0)} = O(\lambda)$ with $\alpha_0$ in the interval $1<\alpha_0 <1.5$.
Variation of $\left\{b_i \right\}$ leads to the variation of $n_{1}^{(0)}$ and $\alpha_0$. Hence to simplify the
qualitative analysis we directly vary $n_{1}^{(0)}$ and $\alpha_0$,  keeping $\left\{b_i \right\}$ fixed; we analyze
the sign of $D$ and $g_{1n}+g_{2\alpha}$ in the according domain, $0.1 \lambda \leq n_{1}^{(0)} \leq 10 \lambda$ and
$1<\alpha_0 <1.5$ for different $\lambda$. The results are shown in Fig.~\ref{fig:Zones}.

Figure \ref{fig:Zones} demonstrates that for $\lambda=0.0001$ there is a large area in the domain
of the $\left(\alpha_0,n_{1}^{(0)}\right)$ plane where steady oscillations may be observed. These may be either linearly stable
oscillations or the growing ones, stabilized by non-linear terms. For relatively large $\lambda
=0.1$, the steady oscillations may arise only in a tiny part of the $\left(\alpha_0,n_{1}^{(0)}\right)$ plane. This
corresponds to the kinetic regimes observed for the full set of aggregation-fragmentation equations: the emergence of
the oscillations for small values of $\lambda$ and their absence for large $\lambda$.

We also note that the average slope of the concentration distribution, $\alpha_0$, is located
within the interval $1.05 \leq \alpha_0 \leq 1.45$, see Fig.~\ref{fig:Zones}, with the median
of $1.25$. This is consistent with the slope of the averaged over oscillations distribution $\left<
n_k\right>_{\rm osc}$ depicted in Fig.~\ref{pic:Oscila1}.

\subsection{Concentration oscillations in thermodynamically closed systems}

As noted above, the time-independent rates $K_{ij}$ and $F_{ij}= \lambda K_{ij}$ imply a steady supply of
energy. This follows generally from the second law of thermodynamics, which excludes steady cyclic processes without
energy supply and may be illustrated for a particular microscopic mechanism of a ballistic aggregation and shattering; this happens in planetary rings or atmospheric processes. Indeed, the conservation of momentum of coalescing particles dictates a withdrawal of a part of their kinetic energy, associated with the relative motion. This energy is transmitted either to the  internal degrees of freedom of the particles (as in planetary rings), or to the surrounding gas (as in atmospheric processes). Similarly, the total kinetic energy of fragments  is smaller than the initial kinetic energy of the colliding aggregates, since part of the  energy is spent to break inter-fragment bonds. Hence, both aggregation and fragmentation processes lead to a gradual reduction of the total kinetic energy of the system. This causes a slowdown of the both processes, and the respective decrease of the rate coefficients. Here we consider thermodynamically closed systems, where the energy supply is lacking. Namely, we consider systems of particles undergoing ballistic aggregation and fragmentation. We chose such systems since the corresponding microscopic rates $K_{ij}$ and $F_{ij}$ are available \cite{Brill2009,PNAS,BFP2018}, see the Appendix.

Physically, the decay of kinetic energy causes a permanent decrease of collision frequency and the decrease of the kinetic rates. Moreover, the fragmentation rates additionally decrease, since the fraction of fast particles, which cause shattering, also drops down. Referring for detail to the Appendix, we present here the equations for
aggregation and shattering for thermodynamically closed systems: 
\begin{eqnarray}
\label{eq:n1ThCl} 
\frac{dn_1}{dt} & = & \frac{\lambda}{2}\sum_{i \geq  2}\sum_{j\geq 2} (i+j) K_{i,j} n_i n_j -n_1 \sum_{i\geq 1} K_{1,i} n_i \nonumber \\
                         &+& \lambda n_1 \sum_{j\geq 2} j K_{1,j} n_j 
\end{eqnarray}
while for heavier clusters, $k\geq 2$, we have
\begin{equation}
\label{eq:nkThCl}
 \frac{dn_k}{dt} = \frac{1}{2} \sum\limits_{i=1}^{k-1} K_{i,k-i} n_i n_{k-i} \!- \!(1 +
\lambda) n_k \sum\limits_{i\geq 1}\!K_{k,i} n_i 
\end{equation}
\begin{equation}
\label{eq:lambThCl}
\lambda = \exp\left[ -A (1+B t)^{b} \right]
\end{equation}
$K_{ij}$ appearing in Eqs.~\eqref{eq:n1ThCl}--\eqref{eq:nkThCl} are defined by \eqref{eq:Cijgen}. Time is measured in collision units, $\tau_c^{-1} = 2 \sqrt{2 \pi}\sigma^2 n_{1,0} \sqrt{T(t)/m}$, where $\sigma$, $m$ and $n_{1,0}$ are respectively the diameter, mass and initial concentration of monomers and $T(t)$ is the characteristic
temperature. The concentration of aggregates is measured in units of the initial  concentration of monomers, $n_{1,0}$.
The constants $A$, $B$ and $b$ in Eq. \eqref{eq:lambThCl} are  expressed respectively in terms of the characteristic
fragmentation energy and temperature decay rate (see Appendix for the detail).

\begin{figure}[h!]
\begin{center}
\includegraphics[scale=0.65]{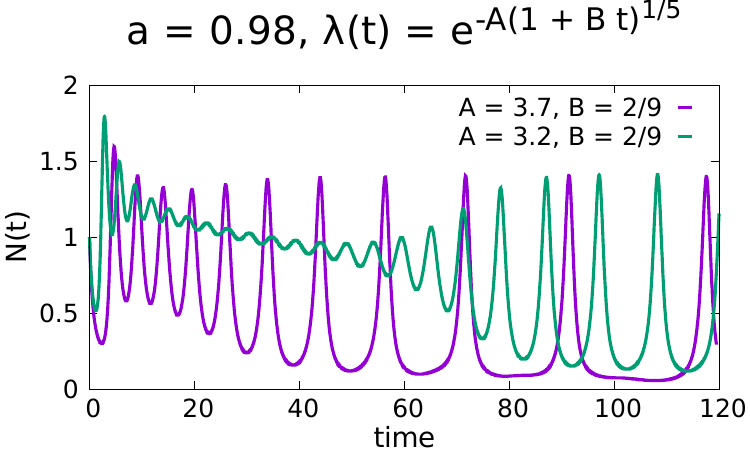}
\caption{ Concentration oscillation in thermodynamically closed systems with aggregation and shattering. Time is
measured in collision units, which keeps the aggregation rates $K_{i,j}= (i/j)^a +(j/i)^a$ with $a=0.98$, steady, but
corresponds to slowing down in the laboratory time. The shattering coefficient $\lambda =\lambda (t)$ decreases with
time according to Eq. \eqref{eq:lambThCl} with $b=0.2$ and different coefficients $A$ and $B$. The persistent
oscillations are clearly visible.  } 
\label{fig:lambThcl}
\end{center}
\end{figure}

The results of numerical solution of Eqs. \eqref{eq:n1ThCl}--\eqref{eq:lambThCl} are presented in Fig.~\ref{fig:lambThcl}. As may be seen from the figure the persistent concentration oscillations emerge in thermodynamically closed systems. Since the time for the depicted oscillations is measured in the collision units, one concludes that the period of these oscillations steadily increases with time; this is also visible in the collision time scale. As one can see from Fig.~\ref{fig:lambThcl}, in thermodynamically closed systems there exist a regime, when the oscillations first decay and then again grow.

\section{Conclusions}
\label{sec:summary}

We have studied numerically and analytically a class of aggregation-fragmentation models. Mathematically, the problem is described by an infinite set of Smoluchowski-like equations with the homogeneous aggregation and fragmentation kernels which respectively read $K_{i,j} = i^{\nu}j^{\mu} + j^{\nu}i^{\mu}$ and $F_{i,j} =\lambda K_{i,j}$, where the parameter $\lambda$ quantifies the intensity of fragmentation. We consider the case of a complete decomposition (shattering) of colliding aggregates into monomers. This model and a similar model, with a source of monomers and evaporation (instead of shattering) of large clusters has been studied recently in
\cite{connaughton2016universality,colmPRE2018}. For the kernels with $\theta=\nu-\mu <1$ we obtain an analytical solution for the steady-state size distribution of the aggregates $n_k$ and confirm numerically the relaxation of the size distribution to this steady-state form.  For kernels with $\theta=\nu-\mu >1$, we observe that the dynamic of the system dramatically depends on the value of the fragmentation constant $\lambda$. While for $\lambda < \lambda_c$ the
system relaxes to a steady-state through damped oscillations of concentrations, for $\lambda \geq \lambda_c$ no
steady-state distribution of the cluster size has been detected.

The emergence of stable oscillations in a closed system of aggregating and fragmenting particles, that lacks any sinks and sources of mass, and formally corresponds to an infinite number of species, is new and surprising. Persistent oscillations have been detected not only for systems, closed with respect to the total mass, but also for
\emph{thermodynamically} closed systems, when the notion ``thermodynamics"  is meaningful.  In Ref.~\cite{Colm} stable oscillations have been detected numerically for
Smoluchowski equations for an open system of reversibly aggregating particles (without fragmentation) with a source of monomers and sink of large clusters, which makes the system finite. For a small closed system comprising monomers, dimers, trimers and exited monomers, stable oscillations of concentrations have been also reported \cite{Gorban}. Similarly, steady chemical oscillations have been found in a simple dimerization model (see e.g.~\cite{dimer} and references therein).

Our findings may  help to understand various phenomena observed in the systems with aggregation and fragmentation, in particular the periodic formation and destruction of clumps in F Ring of Saturn \cite{FRing}, where particles of different mass suffer aggregative and disruptive impacts, presumably  under the
 mass conservation condition. A complete understanding of this phenomenon is presently lacking.

\appendix
\section{}

General expressions for the aggregation and fragmentation rates for a system of ballistically moving particles
(molecules, macroscopic grains, etc.) that suffer pairwise collisions, have been reported in 
\cite{Brill2009,PNAS}. These rates read
\begin{equation}
 \label{eq:KF-nu}
 \begin{split}
K_{ij} &= \nu_{ij} \left[ 1- \left(1+ B_{ij}\tilde{E}_{\rm agg}\right)e^{-B_{ij}\tilde{E}_{\rm agg}} \right] \\
F_{ij} &= \nu_{ij} e^{-B_{ij}E_{\rm frag}}, \\
\nu_{ij} &= 2 \sqrt{2 \pi} \sigma_{ij}^2 \sqrt{\frac{T_i}{m_i} +\frac{T_j}{m_j} } \\
B_{ij} &= \frac{m_i^{-1} +m_j^{-1}}{T_i/m_i +T_j/m_j} ,
 \end{split}
\end{equation}
in present notations. Here $T_i$ are partial temperatures of aggregates of size $i$, and mass $m_i=m_1 i$  ($m_1$ is the mass of monomer),
that characterizes the average kinetic energy of such aggregates \cite{Brill2009,PNAS,BFP2018}. $E_{\rm agg}$ and
$E_{\rm frag}$ are respectively the aggregation and fragmentation energy, $\tilde{E}_{\rm agg}= E_{\rm
agg}/\varepsilon^2$, where $\varepsilon$ is the coefficient of normal restitution that characterizes dissipative losses
in the impacts \cite{BFP2018}. $\sigma_{ij}= \sigma_1 \l(i^{1/D} +j^{1/D} \r)$, where $\sigma_1$ is monomer diameter
and $D$ is the dimension of the aggregates, which may be fractal. Based on the results of Ref. \cite{bodrova2014}, we
assume that partial temperatures scale as $T_i(t) =T(t) i^{\gamma}$, where $T(t)$ is the characteristic temperature of
the gas mixture. (It may be shown that after a short relaxation time the rate of change of temperatures of all species
is the same, $T_i^{-1} dT_i/dt = T^{-1} dT/dt$, \cite{bodrova2014}).

It is convenient to recast the above kinetic coefficients into the fform
\begin{equation}
 \label{eq:KijFij}
 \begin{split}
K_{ij} &= \tau_0^{-1} (T/T_0)^{1/2} n_{1,0}^{-1} \tilde{K}_{ij} \\
\tilde{K}_{ij}&= \l(i^{1/D} +j^{1/D} \r)^2 \l(i^{\gamma-1} + j^{\gamma - 1} \r)^{1/2} \\
F_{ij} &= \lambda K_{ij} ,\qq \lambda  \approx \exp \l[ -A \, (T_0/T) \r]
 \end{split}
\end{equation}
where $n_{1,0}$ is the initial concentration of monomers, which we will use as a unit of concentration, $T_0$ is the
initial characteristic  temperature, and
\be 
\lb{eq:tau0} 
\tau_0^{-1} = 2 \sqrt{2 \pi} \sigma_0^2 n_{1,0} (T_0/m_1)^{1/2}  
\ee
gives the initial characteristic collision frequency. The quantity 
\begin{equation*}
A = \frac{E_{\rm frag}}{T_0}\left\langle \frac{i^{-1} + j^{-1}}{ i^{\gamma-1} + j^{\gamma-1}}\right \rangle
\end{equation*}
appearing in \eqref{eq:KijFij} is the effective average ratio of fragmentation and kinetic energy. We also assume for simplicity that the aggregation energy is large, so that $B_{ij}\tilde{E}_{\rm agg} \gg 1$.

The dimensionless kernels $\tilde{K}_{ij}$ are obtained by a straightforward solution of the Boltzmann equation
\cite{Brill2009,PNAS,BFP2018}. Here we apply a standard simplification \cite{Krapivsky,Leyvraz} for these
kernels, which allows analytical analysis. It is based on the observation that the main properties of the solutions to
the Smoluchowski equations depend on two indices $\beta$ and $\beta_1$, characterizing the kinetic rates $K_{ij}$. The
first index quantifies the homogeneity degree of a  kernel, and the second one the size dependence at  the
maximal size asymmetry. Namely,
$$
\tilde{K}_{ai, aj} \sim a^{\beta} \tilde {K}_{ij}; \qq \tilde{K}_{1,j} \sim j^{\beta_1} \qq {\rm for } \qq i,j \gg 1.
$$
For the kernels 
\begin{equation*}
\tilde{K}_{ij}=i^{\nu}j^{\mu} + i^{\mu} j^{\nu}
\end{equation*}
introduced in Eq.~\eqref{eq:Cijgen}, one obtains $\beta=\nu+\mu$ and $\beta_1 = {\rm max}(\nu, \mu)$. Therefore Eqs.~\eqref{eq:KijFij} show that $\nu$ and $\mu$ are related to the physical parameters $D$ and $\gamma$ via relations $(\nu, \mu)= \l(2/D, (\gamma-1)/2 \r)$ for $\gamma<1$ and $(\nu, \mu)= \l( 2/D +(\gamma-1)/2, 0 \r)$ for $\gamma>1$.

Next, we derive the equation for the characteristic temperature $T(t)$. This may be done using the approach of Ref.~\cite{BFP2018}, which yields
\be 
\lb{eq:eqforT} 
\frac{d}{dt} N T = -\sum_{ij} Q_{ij}(T) n_in_j + \sum_{i}\Gamma_i n_i.  
\ee
Here $Q_{ij}(T)$ are temperature-dependent rate coefficients and $\Gamma_i$ describes the energy input to the system due
to the interaction of the aggregates of size $i$ with the external sources of energy (see also \cite{bodrova2014}).
Here we do not need explicit expressions for these quantities. We just state  that the presence of the energy sources
$\Gamma_i$ in  Eq.~\eqref{eq:eqforT} yields the solutions with a constant temperature $T=\rm const.$, corresponding to
the systems with time-independent rates $K_{ij}$ and $F_{ij}$. For thermodynamically closed systems, the temperature
commonly decreases with time (see the discussion in Ref. \cite{BFP2018}).

The solutions of the coupled Smoluchowski-like  equations \eqref{eq:KijFij} and \eqref{eq:eqforT} for concentrations
and temperature is beyond the scope of the present study. For the qualitative analysis, we assume a power-law decay of
the characteristic temperature with time,  $T = T_0(1+t/\tau_0)^{-\delta }$; such assumption is justified by the
results of Ref. \cite{BFP2018}. The value of $\delta$ depends on the parameters of the system and may vary in a wide
interval \cite{BFP2018}. Using the collision frequency, $\tau_c^{-1}(t)= \tau_0^{-1} \l[T(t)/T_0\r]^{1/2}$ at the
current time $t$, we introduce a  new dimensionless time $\tilde{t}$, measured in collision units. It is related to the
laboratory time as $\tau_c^{-1}(t)dt = d\tilde{t} $. The dependence of temperature on the new time then reads
\be 
\lb{eq:Tnewt} 
T/T_0 = \l[1 + B \tilde{t} \,\r]^{-b},  
\ee
where $b=2\delta /(2-\delta )$ and $B$ is a constant. The kinetic rates may be expressed in terms of the collision-based time $\tilde t$:
\beq 
\lb{eq:Kijnewt}
K_{ij} \eq \tau_c^{-1}(\tilde t\,) n_{1,0}^{-1} \tilde{K}_{ij}  \\ 
\lb{eq:lamnewt}
\lambda \eq
\exp\l[-A (1+B\tilde{t}\,)^b  \r]. 
\eeq
In our simulations we choose $b= 0.2$ (which corresponds to $\delta =10/11$).

Substituting the rates $K_{ij}$ and $\lambda$ from Eqs.~\eqref{eq:Kijnewt} and \eqref{eq:lamnewt} into Eqs.~\eqref{eq:n1} and \eqref{eq:nk}, we arrive at 
Eqs.~\eqref{eq:n1ThCl}--\eqref{eq:lambThCl}, where time is measured in the collision units and concentrations in the units of initial concentration of monomers. For simplicity we use in these equations the same notations as in Eqs.~\eqref{eq:n1} and \eqref{eq:nk}.

\bibliography{oscillations1}

\begin{thebibliography}{42}%
\makeatletter
\providecommand \@ifxundefined [1]{%
 \@ifx{#1\undefined}
}%
\providecommand \@ifnum [1]{%
 \ifnum #1\expandafter \@firstoftwo
 \else \expandafter \@secondoftwo
 \fi
}%
\providecommand \@ifx [1]{%
 \ifx #1\expandafter \@firstoftwo
 \else \expandafter \@secondoftwo
 \fi
}%
\providecommand \natexlab [1]{#1}%
\providecommand \enquote  [1]{``#1''}%
\providecommand \bibnamefont  [1]{#1}%
\providecommand \bibfnamefont [1]{#1}%
\providecommand \citenamefont [1]{#1}%
\providecommand \href@noop [0]{\@secondoftwo}%
\providecommand \href [0]{\begingroup \@sanitize@url \@href}%
\providecommand \@href[1]{\@@startlink{#1}\@@href}%
\providecommand \@@href[1]{\endgroup#1\@@endlink}%
\providecommand \@sanitize@url [0]{\catcode `\\12\catcode `\$12\catcode
  `\&12\catcode `\#12\catcode `\^12\catcode `\_12\catcode `\%12\relax}%
\providecommand \@@startlink[1]{}%
\providecommand \@@endlink[0]{}%
\providecommand \url  [0]{\begingroup\@sanitize@url \@url }%
\providecommand \@url [1]{\endgroup\@href {#1}{\urlprefix }}%
\providecommand \urlprefix  [0]{URL }%
\providecommand \Eprint [0]{\href }%
\providecommand \doibase [0]{http://dx.doi.org/}%
\providecommand \selectlanguage [0]{\@gobble}%
\providecommand \bibinfo  [0]{\@secondoftwo}%
\providecommand \bibfield  [0]{\@secondoftwo}%
\providecommand \translation [1]{[#1]}%
\providecommand \BibitemOpen [0]{}%
\providecommand \bibitemStop [0]{}%
\providecommand \bibitemNoStop [0]{.\EOS\space}%
\providecommand \EOS [0]{\spacefactor3000\relax}%
\providecommand \BibitemShut  [1]{\csname bibitem#1\endcsname}%
\let\auto@bib@innerbib\@empty
\bibitem [{\citenamefont {Krapivsky}\ \emph {et~al.}(2010)\citenamefont
  {Krapivsky}, \citenamefont {Redner},\ and\ \citenamefont
  {Ben-Naim}}]{Krapivsky}%
  \BibitemOpen
  \bibfield  {author} {\bibinfo {author} {\bibfnamefont {P.~L.}\ \bibnamefont
  {Krapivsky}}, \bibinfo {author} {\bibfnamefont {S.}~\bibnamefont {Redner}}, \
  and\ \bibinfo {author} {\bibfnamefont {E.}~\bibnamefont {Ben-Naim}},\ }\href
  {https://doi.org/10.1017/CBO9780511780516} {\emph {\bibinfo {title} {A
  Kinetic View of Statistical Physics}}}\ (\bibinfo  {publisher} {Cambridge
  University Press},\ \bibinfo {address} {Cambridge, UK},\ \bibinfo {year}
  {2010})\BibitemShut {NoStop}%
\bibitem [{\citenamefont {Leyvraz}(2003)}]{Leyvraz}%
  \BibitemOpen
  \bibfield  {author} {\bibinfo {author} {\bibfnamefont {F.}~\bibnamefont
  {Leyvraz}},\ }\bibfield  {title} {\enquote {\bibinfo {title} {Scaling theory
  and exactly solved models in the kinetics of irreversible aggregation},}\
  }\href {https://doi.org/10.1016/S0370-1573(03)00241-2} {\bibfield  {journal}
  {\bibinfo  {journal} {Phys. Reports}\ }\textbf {\bibinfo {volume} {383}},\
  \bibinfo {pages} {95--212} (\bibinfo {year} {2003})}\BibitemShut {NoStop}%
\bibitem [{\citenamefont {Poeschel}\ \emph {et~al.}(2003)\citenamefont
  {Poeschel}, \citenamefont {Brilliantov},\ and\ \citenamefont
  {Frommel}}]{prions}%
  \BibitemOpen
  \bibfield  {author} {\bibinfo {author} {\bibfnamefont {T.}~\bibnamefont
  {Poeschel}}, \bibinfo {author} {\bibfnamefont {N.~V.}\ \bibnamefont
  {Brilliantov}}, \ and\ \bibinfo {author} {\bibfnamefont {C.}~\bibnamefont
  {Frommel}},\ }\bibfield  {title} {\enquote {\bibinfo {title} {Kinetics of
  prion growth},}\ }\href {\doibase
  https://doi.org/10.1016/S0006-3495(03)74767-5} {\bibfield  {journal}
  {\bibinfo  {journal} {Biophys. J.}\ }\textbf {\bibinfo {volume} {85}},\
  \bibinfo {pages} {3460--3474} (\bibinfo {year} {2003})}\BibitemShut {NoStop}%
\bibitem [{\citenamefont {Srivastava}(1982)}]{Srivastava1982}%
  \BibitemOpen
  \bibfield  {author} {\bibinfo {author} {\bibfnamefont {R.~C.}\ \bibnamefont
  {Srivastava}},\ }\bibfield  {title} {\enquote {\bibinfo {title} {A simple
  model of particle coalescence and breakup},}\ }\href
  {https://journals.ametsoc.org/view/journals/atsc/39/6/1520-0469_1982_039_1317_asmopc_2_0_co_2.xml}
  {\bibfield  {journal} {\bibinfo  {journal} {J. Atmos. Sci.}\ }\textbf
  {\bibinfo {volume} {39}},\ \bibinfo {pages} {1317--1321} (\bibinfo {year}
  {1982})}\BibitemShut {NoStop}%
\bibitem [{\citenamefont {Grant}(1994)}]{Grant}%
  \BibitemOpen
  \bibfield  {author} {\bibinfo {author} {\bibfnamefont {S.~B.}\ \bibnamefont
  {Grant}},\ }\bibfield  {title} {\enquote {\bibinfo {title} {Virus coagulation
  in aqueous environments},}\ }\href {https://doi.org/10.1021/es00054a026}
  {\bibfield  {journal} {\bibinfo  {journal} {Environ. Sci. Technol.}\ }\textbf
  {\bibinfo {volume} {28}},\ \bibinfo {pages} {928--933} (\bibinfo {year}
  {1994})}\BibitemShut {NoStop}%
\bibitem [{\citenamefont {Niwa}(1998)}]{Niwa}%
  \BibitemOpen
  \bibfield  {author} {\bibinfo {author} {\bibfnamefont {H.-S.}\ \bibnamefont
  {Niwa}},\ }\bibfield  {title} {\enquote {\bibinfo {title} {School size
  statistics of fish},}\ }\href {\doibase
  https://doi.org/10.1006/jtbi.1998.0801} {\bibfield  {journal} {\bibinfo
  {journal} {J. Theor. Biol.}\ }\textbf {\bibinfo {volume} {195}},\ \bibinfo
  {pages} {351--361} (\bibinfo {year} {1998})}\BibitemShut {NoStop}%
\bibitem [{\citenamefont {Miura}\ \emph {et~al.}(2012)\citenamefont {Miura},
  \citenamefont {Takayasu},\ and\ \citenamefont {Takayasu}}]{Takayasu}%
  \BibitemOpen
  \bibfield  {author} {\bibinfo {author} {\bibfnamefont {W.}~\bibnamefont
  {Miura}}, \bibinfo {author} {\bibfnamefont {H.}~\bibnamefont {Takayasu}}, \
  and\ \bibinfo {author} {\bibfnamefont {M.}~\bibnamefont {Takayasu}},\
  }\bibfield  {title} {\enquote {\bibinfo {title} {Effect of coagulation of
  nodes in an evolving complex network},}\ }\href {\doibase
  10.1103/PhysRevLett.108.168701} {\bibfield  {journal} {\bibinfo  {journal}
  {Phys. Rev. Lett.}\ }\textbf {\bibinfo {volume} {108}},\ \bibinfo {pages}
  {168701} (\bibinfo {year} {2012})}\BibitemShut {NoStop}%
\bibitem [{\citenamefont {Dorogovtsev}\ and\ \citenamefont
  {Mendes}(2003)}]{Dorogov}%
  \BibitemOpen
  \bibfield  {author} {\bibinfo {author} {\bibfnamefont {S.~N.}\ \bibnamefont
  {Dorogovtsev}}\ and\ \bibinfo {author} {\bibfnamefont {J.~F.~F.}\
  \bibnamefont {Mendes}},\ }\href {\doibase
  https://doi.org/10.1093/acprof:oso/9780198515906.001.0001} {\emph {\bibinfo
  {title} {Evolution of networks: From biological nets to the Internet and
  WWW}}}\ (\bibinfo  {publisher} {Oxford University Press},\ \bibinfo {address}
  {Oxford, UK},\ \bibinfo {year} {2003})\BibitemShut {NoStop}%
\bibitem [{\citenamefont {Zakharov}\ \emph {et~al.}(2012)\citenamefont
  {Zakharov}, \citenamefont {L'vov},\ and\ \citenamefont
  {Falkovich}}]{Turbulence}%
  \BibitemOpen
  \bibfield  {author} {\bibinfo {author} {\bibfnamefont {V.~E.}\ \bibnamefont
  {Zakharov}}, \bibinfo {author} {\bibfnamefont {V.~S.}\ \bibnamefont {L'vov}},
  \ and\ \bibinfo {author} {\bibfnamefont {G.}~\bibnamefont {Falkovich}},\
  }\href@noop {} {\emph {\bibinfo {title} {Kolmogorov Spectra of Turbulence I:
  Wave Turbulence}}}\ (\bibinfo  {publisher} {Springer},\ \bibinfo {address}
  {New York, NY},\ \bibinfo {year} {2012})\BibitemShut {NoStop}%
\bibitem [{\citenamefont {Brilliantov}\ \emph {et~al.}(2015)\citenamefont
  {Brilliantov}, \citenamefont {Krapivsky}, \citenamefont {Bodrova},
  \citenamefont {Spahn}, \citenamefont {Hayakawa}, \citenamefont {Stadnichuk},\
  and\ \citenamefont {Schmidt}}]{PNAS}%
  \BibitemOpen
  \bibfield  {author} {\bibinfo {author} {\bibfnamefont {N.~V.}\ \bibnamefont
  {Brilliantov}}, \bibinfo {author} {\bibfnamefont {P.~L.}\ \bibnamefont
  {Krapivsky}}, \bibinfo {author} {\bibfnamefont {A.}~\bibnamefont {Bodrova}},
  \bibinfo {author} {\bibfnamefont {F.}~\bibnamefont {Spahn}}, \bibinfo
  {author} {\bibfnamefont {H.}~\bibnamefont {Hayakawa}}, \bibinfo {author}
  {\bibfnamefont {V.}~\bibnamefont {Stadnichuk}}, \ and\ \bibinfo {author}
  {\bibfnamefont {J.}~\bibnamefont {Schmidt}},\ }\bibfield  {title} {\enquote
  {\bibinfo {title} {Size distribution of particles in {S}aturn's rings from
  aggregation and fragmentation},}\ }\href
  {https://doi.org/10.1073/pnas.1503957112} {\bibfield  {journal} {\bibinfo
  {journal} {PNAS}\ }\textbf {\bibinfo {volume} {112}},\ \bibinfo {pages}
  {9536--9541} (\bibinfo {year} {2015})}\BibitemShut {NoStop}%
\bibitem [{\citenamefont {Stadnichuk}\ \emph {et~al.}(2015)\citenamefont
  {Stadnichuk}, \citenamefont {Bodrova},\ and\ \citenamefont
  {Brilliantov}}]{stadnichuk2015smoluchowski}%
  \BibitemOpen
  \bibfield  {author} {\bibinfo {author} {\bibfnamefont {V.}~\bibnamefont
  {Stadnichuk}}, \bibinfo {author} {\bibfnamefont {A.}~\bibnamefont {Bodrova}},
  \ and\ \bibinfo {author} {\bibfnamefont {N.~V.}\ \bibnamefont
  {Brilliantov}},\ }\bibfield  {title} {\enquote {\bibinfo {title}
  {Smoluchowski aggregation--fragmentation equations: Fast numerical method to
  find steady-state solutions},}\ }\href
  {https://doi.org/10.1142/S0217979215502082} {\bibfield  {journal} {\bibinfo
  {journal} {Int. J. Mod. Phys. B}\ }\textbf {\bibinfo {volume} {29}},\
  \bibinfo {pages} {1550208} (\bibinfo {year} {2015})}\BibitemShut {NoStop}%
\bibitem [{\citenamefont {Cuzzi}\ \emph {et~al.}(2010)\citenamefont {Cuzzi},
  \citenamefont {Burns}, \citenamefont {Charnoz}, \citenamefont {Clark},
  \citenamefont {Colwell}, \citenamefont {Dones}, \citenamefont {Esposito},
  \citenamefont {Filacchione}, \citenamefont {French}, \citenamefont {Hedman}
  \emph {et~al.}}]{Cuzzi}%
  \BibitemOpen
  \bibfield  {author} {\bibinfo {author} {\bibfnamefont {J.~N.}\ \bibnamefont
  {Cuzzi}}, \bibinfo {author} {\bibfnamefont {J.~A.}\ \bibnamefont {Burns}},
  \bibinfo {author} {\bibfnamefont {S.}~\bibnamefont {Charnoz}}, \bibinfo
  {author} {\bibfnamefont {R.~N.}\ \bibnamefont {Clark}}, \bibinfo {author}
  {\bibfnamefont {J.~E.}\ \bibnamefont {Colwell}}, \bibinfo {author}
  {\bibfnamefont {L.}~\bibnamefont {Dones}}, \bibinfo {author} {\bibfnamefont
  {L.~W.}\ \bibnamefont {Esposito}}, \bibinfo {author} {\bibfnamefont
  {G.}~\bibnamefont {Filacchione}}, \bibinfo {author} {\bibfnamefont {R.~G.}\
  \bibnamefont {French}}, \bibinfo {author} {\bibfnamefont {M.~M.}\
  \bibnamefont {Hedman}},  \emph {et~al.},\ }\bibfield  {title} {\enquote
  {\bibinfo {title} {An evolving view of {S}aturn's dynamic rings},}\ }\href
  {\doibase 10.1126/science.11791} {\bibfield  {journal} {\bibinfo  {journal}
  {Science}\ }\textbf {\bibinfo {volume} {327}},\ \bibinfo {pages} {1470--1475}
  (\bibinfo {year} {2010})}\BibitemShut {NoStop}%
\bibitem [{\citenamefont {Brilliantov}\ \emph {et~al.}(2009)\citenamefont
  {Brilliantov}, \citenamefont {Bodrova},\ and\ \citenamefont
  {Krapivsky}}]{Brill2009}%
  \BibitemOpen
  \bibfield  {author} {\bibinfo {author} {\bibfnamefont {N.~V.}\ \bibnamefont
  {Brilliantov}}, \bibinfo {author} {\bibfnamefont {A.}~\bibnamefont
  {Bodrova}}, \ and\ \bibinfo {author} {\bibfnamefont {P.~L.}\ \bibnamefont
  {Krapivsky}},\ }\bibfield  {title} {\enquote {\bibinfo {title} {A model of
  ballistic aggregation and fragmentation},}\ }\href {\doibase
  10.1088/1742-5468/2009/06/P06011} {\bibfield  {journal} {\bibinfo  {journal}
  {J. Stat. Mech.}\ }\textbf {\bibinfo {volume} {2009}},\ \bibinfo {pages}
  {P06011} (\bibinfo {year} {2009})}\BibitemShut {NoStop}%
\bibitem [{\citenamefont {Esposito}(2006)}]{Esposito}%
  \BibitemOpen
  \bibfield  {author} {\bibinfo {author} {\bibfnamefont {L.}~\bibnamefont
  {Esposito}},\ }\href {https://doi.org/10.1017/CBO9781139236966} {\emph
  {\bibinfo {title} {Planetary rings}}}\ (\bibinfo  {publisher} {Cambridge
  University Press},\ \bibinfo {address} {Cambridge, UK},\ \bibinfo {year}
  {2006})\BibitemShut {NoStop}%
\bibitem [{\citenamefont {Krapivsky}\ \emph {et~al.}(2017)\citenamefont
  {Krapivsky}, \citenamefont {Otieno},\ and\ \citenamefont
  {Brilliantov}}]{KOB2017}%
  \BibitemOpen
  \bibfield  {author} {\bibinfo {author} {\bibfnamefont {P.~L.}\ \bibnamefont
  {Krapivsky}}, \bibinfo {author} {\bibfnamefont {W.}~\bibnamefont {Otieno}}, \
  and\ \bibinfo {author} {\bibfnamefont {N.~V.}\ \bibnamefont {Brilliantov}},\
  }\bibfield  {title} {\enquote {\bibinfo {title} {Phase transitions in systems
  with aggregation and shattering},}\ }\href {\doibase
  10.1103/PhysRevE.96.042138} {\bibfield  {journal} {\bibinfo  {journal} {Phys.
  Rev. E}\ }\textbf {\bibinfo {volume} {96}},\ \bibinfo {pages} {042138}
  (\bibinfo {year} {2017})}\BibitemShut {NoStop}%
\bibitem [{\citenamefont {Brilliantov}\ \emph {et~al.}(2018)\citenamefont
  {Brilliantov}, \citenamefont {Formella},\ and\ \citenamefont
  {Poeschel}}]{BFP2018}%
  \BibitemOpen
  \bibfield  {author} {\bibinfo {author} {\bibfnamefont {N.~V.}\ \bibnamefont
  {Brilliantov}}, \bibinfo {author} {\bibfnamefont {A.}~\bibnamefont
  {Formella}}, \ and\ \bibinfo {author} {\bibfnamefont {T.}~\bibnamefont
  {Poeschel}},\ }\bibfield  {title} {\enquote {\bibinfo {title} {Increasing
  temperature of cooling granular gases},}\ }\href
  {https://doi.org/10.1038/s41467-017-02803-7} {\bibfield  {journal} {\bibinfo
  {journal} {Nature Commun.}\ }\textbf {\bibinfo {volume} {9}},\ \bibinfo
  {pages} {797} (\bibinfo {year} {2018})}\BibitemShut {NoStop}%
\bibitem [{\citenamefont {Connaughton}\ \emph {et~al.}(2017)\citenamefont
  {Connaughton}, \citenamefont {Dutta}, \citenamefont {Rajesh},\ and\
  \citenamefont {Zaboronski}}]{connaughton2016universality}%
  \BibitemOpen
  \bibfield  {author} {\bibinfo {author} {\bibfnamefont {C.}~\bibnamefont
  {Connaughton}}, \bibinfo {author} {\bibfnamefont {A.}~\bibnamefont {Dutta}},
  \bibinfo {author} {\bibfnamefont {R.}~\bibnamefont {Rajesh}}, \ and\ \bibinfo
  {author} {\bibfnamefont {O.}~\bibnamefont {Zaboronski}},\ }\bibfield  {title}
  {\enquote {\bibinfo {title} {Universality properties of steady driven
  coagulation with collisional evaporation},}\ }\href {\doibase
  10.1209/0295-5075/117/10002} {\bibfield  {journal} {\bibinfo  {journal}
  {Europhys. Lett.}\ }\textbf {\bibinfo {volume} {117}},\ \bibinfo {pages}
  {10002} (\bibinfo {year} {2017})}\BibitemShut {NoStop}%
\bibitem [{\citenamefont {Connaughton}\ \emph {et~al.}(2018)\citenamefont
  {Connaughton}, \citenamefont {Dutta}, \citenamefont {Rajesh}, \citenamefont
  {Siddharth},\ and\ \citenamefont {Zaboronski}}]{colmPRE2018}%
  \BibitemOpen
  \bibfield  {author} {\bibinfo {author} {\bibfnamefont {C.}~\bibnamefont
  {Connaughton}}, \bibinfo {author} {\bibfnamefont {A.}~\bibnamefont {Dutta}},
  \bibinfo {author} {\bibfnamefont {R.}~\bibnamefont {Rajesh}}, \bibinfo
  {author} {\bibfnamefont {N.}~\bibnamefont {Siddharth}}, \ and\ \bibinfo
  {author} {\bibfnamefont {O.}~\bibnamefont {Zaboronski}},\ }\bibfield  {title}
  {\enquote {\bibinfo {title} {Stationary mass distribution and nonlocality in
  models of coalescence and shattering},}\ }\href {\doibase
  10.1103/PhysRevE.97.022137} {\bibfield  {journal} {\bibinfo  {journal} {Phys.
  Rev. E}\ }\textbf {\bibinfo {volume} {97}},\ \bibinfo {pages} {022137}
  (\bibinfo {year} {2018})}\BibitemShut {NoStop}%
\bibitem [{\citenamefont {Hendriks}\ \emph {et~al.}(1983)\citenamefont
  {Hendriks}, \citenamefont {Ernst},\ and\ \citenamefont {Ziff}}]{Ernst1983}%
  \BibitemOpen
  \bibfield  {author} {\bibinfo {author} {\bibfnamefont {E.~M.}\ \bibnamefont
  {Hendriks}}, \bibinfo {author} {\bibfnamefont {M.~H.}\ \bibnamefont {Ernst}},
  \ and\ \bibinfo {author} {\bibfnamefont {R.~M.}\ \bibnamefont {Ziff}},\
  }\bibfield  {title} {\enquote {\bibinfo {title} {Coagulation equations with
  gelation},}\ }\href {https://doi.org/10.1007/BF01019497} {\bibfield
  {journal} {\bibinfo  {journal} {J. Stat. Phys.}\ }\textbf {\bibinfo {volume}
  {31}},\ \bibinfo {pages} {519--563} (\bibinfo {year} {1983})}\BibitemShut
  {NoStop}%
\bibitem [{\citenamefont {van Dongen}(1987)}]{van87}%
  \BibitemOpen
  \bibfield  {author} {\bibinfo {author} {\bibfnamefont {P.~G.~J.}\
  \bibnamefont {van Dongen}},\ }\bibfield  {title} {\enquote {\bibinfo {title}
  {On the possible occurrence of instantaneous gelation in {S}moluchowski's
  coagulation equation},}\ }\href {\doibase 10.1088/0305-4470/20/7/033}
  {\bibfield  {journal} {\bibinfo  {journal} {J. Phys. A}\ }\textbf {\bibinfo
  {volume} {20}},\ \bibinfo {pages} {1889--1904} (\bibinfo {year}
  {1987})}\BibitemShut {NoStop}%
\bibitem [{\citenamefont {Brilliantov}\ and\ \citenamefont
  {Krapivsky}(1991)}]{bk}%
  \BibitemOpen
  \bibfield  {author} {\bibinfo {author} {\bibfnamefont {N.~V.}\ \bibnamefont
  {Brilliantov}}\ and\ \bibinfo {author} {\bibfnamefont {P.~L.}\ \bibnamefont
  {Krapivsky}},\ }\bibfield  {title} {\enquote {\bibinfo {title} {Nonscaling
  and source-induced scaling behaviour in aggregation model of movable monomers
  and immovable clusters},}\ }\href {\doibase 10.1088/0305-4470/24/20/014}
  {\bibfield  {journal} {\bibinfo  {journal} {J. Phys. A}\ }\textbf {\bibinfo
  {volume} {24}},\ \bibinfo {pages} {4787--4803} (\bibinfo {year}
  {1991})}\BibitemShut {NoStop}%
\bibitem [{\citenamefont {Lauren\c{c}ot}(1999)}]{Laurencot1999}%
  \BibitemOpen
  \bibfield  {author} {\bibinfo {author} {\bibfnamefont {Ph.}\ \bibnamefont
  {Lauren\c{c}ot}},\ }\bibfield  {title} {\enquote {\bibinfo {title} {Singular
  behavior of finite approximations to the addition model},}\ }\href {\doibase
  10.1088/0951-7715/12/2/004} {\bibfield  {journal} {\bibinfo  {journal}
  {Nonlinearity}\ }\textbf {\bibinfo {volume} {12}},\ \bibinfo {pages}
  {229--239} (\bibinfo {year} {1999})}\BibitemShut {NoStop}%
\bibitem [{\citenamefont {Malyshkin}\ and\ \citenamefont
  {Goodman}(2001)}]{Malyshkin2001}%
  \BibitemOpen
  \bibfield  {author} {\bibinfo {author} {\bibfnamefont {L.}~\bibnamefont
  {Malyshkin}}\ and\ \bibinfo {author} {\bibfnamefont {J.}~\bibnamefont
  {Goodman}},\ }\bibfield  {title} {\enquote {\bibinfo {title} {The timescale
  of runaway stochastic coagulation},}\ }\href
  {https://doi.org/10.1006/icar.2001.6587} {\bibfield  {journal} {\bibinfo
  {journal} {Icarus}\ }\textbf {\bibinfo {volume} {150}},\ \bibinfo {pages}
  {314--322} (\bibinfo {year} {2001})}\BibitemShut {NoStop}%
\bibitem [{\citenamefont {Krapivsky}\ and\ \citenamefont
  {Connaughton}(2012)}]{ColmPaulJCP2012}%
  \BibitemOpen
  \bibfield  {author} {\bibinfo {author} {\bibfnamefont {P.~L.}\ \bibnamefont
  {Krapivsky}}\ and\ \bibinfo {author} {\bibfnamefont {C.}~\bibnamefont
  {Connaughton}},\ }\bibfield  {title} {\enquote {\bibinfo {title} {Driven
  {B}rownian coagulation of polymers},}\ }\href
  {https://doi.org/10.1063/1.4718833} {\bibfield  {journal} {\bibinfo
  {journal} {J. Chem. Phys.}\ }\textbf {\bibinfo {volume} {136}},\ \bibinfo
  {pages} {204901} (\bibinfo {year} {2012})}\BibitemShut {NoStop}%
\bibitem [{\citenamefont {Ball}\ \emph {et~al.}(2011)\citenamefont {Ball},
  \citenamefont {Connaughton}, \citenamefont {Stein},\ and\ \citenamefont
  {Zaboronski}}]{Colm2011}%
  \BibitemOpen
  \bibfield  {author} {\bibinfo {author} {\bibfnamefont {R.~C.}\ \bibnamefont
  {Ball}}, \bibinfo {author} {\bibfnamefont {C.}~\bibnamefont {Connaughton}},
  \bibinfo {author} {\bibfnamefont {T.~H.~M.}\ \bibnamefont {Stein}}, \ and\
  \bibinfo {author} {\bibfnamefont {O.}~\bibnamefont {Zaboronski}},\ }\bibfield
   {title} {\enquote {\bibinfo {title} {Instantaneous gelation in
  smoluchowski's coagulation equation revisited},}\ }\href {\doibase
  10.1103/PhysRevE.84.011111} {\bibfield  {journal} {\bibinfo  {journal} {Phys.
  Rev. E}\ }\textbf {\bibinfo {volume} {84}},\ \bibinfo {pages} {011111}
  (\bibinfo {year} {2011})}\BibitemShut {NoStop}%
\bibitem [{\citenamefont {Hayakawa}(1987)}]{hayakawa1987irreversible}%
  \BibitemOpen
  \bibfield  {author} {\bibinfo {author} {\bibfnamefont {H.}~\bibnamefont
  {Hayakawa}},\ }\bibfield  {title} {\enquote {\bibinfo {title} {Irreversible
  kinetic coagulations in the presence of a source},}\ }\href {\doibase
  10.1088/0305-4470/20/12/009} {\bibfield  {journal} {\bibinfo  {journal} {J.
  Phys. A}\ }\textbf {\bibinfo {volume} {20}},\ \bibinfo {pages} {L801--L805}
  (\bibinfo {year} {1987})}\BibitemShut {NoStop}%
\bibitem [{\citenamefont {Ball}\ \emph {et~al.}(2012)\citenamefont {Ball},
  \citenamefont {Connaughton}, \citenamefont {Jones}, \citenamefont {Rajesh},\
  and\ \citenamefont {Zaboronski}}]{Colm}%
  \BibitemOpen
  \bibfield  {author} {\bibinfo {author} {\bibfnamefont {R.~C.}\ \bibnamefont
  {Ball}}, \bibinfo {author} {\bibfnamefont {C.}~\bibnamefont {Connaughton}},
  \bibinfo {author} {\bibfnamefont {P.~P.}\ \bibnamefont {Jones}}, \bibinfo
  {author} {\bibfnamefont {R.}~\bibnamefont {Rajesh}}, \ and\ \bibinfo {author}
  {\bibfnamefont {O.}~\bibnamefont {Zaboronski}},\ }\bibfield  {title}
  {\enquote {\bibinfo {title} {Collective oscillations in irreversible
  coagulation driven by monomer inputs and large-cluster outputs},}\ }\href
  {https://link.aps.org/doi/10.1103/PhysRevLett.109.168304} {\bibfield
  {journal} {\bibinfo  {journal} {Phys. Rev. Lett.}\ }\textbf {\bibinfo
  {volume} {109}},\ \bibinfo {pages} {168304} (\bibinfo {year}
  {2012})}\BibitemShut {NoStop}%
\bibitem [{\citenamefont {Bykov}\ and\ \citenamefont {Gorban}(1987)}]{Gorban}%
  \BibitemOpen
  \bibfield  {author} {\bibinfo {author} {\bibfnamefont {V.~I.}\ \bibnamefont
  {Bykov}}\ and\ \bibinfo {author} {\bibfnamefont {A.~N.}\ \bibnamefont
  {Gorban}},\ }\bibfield  {title} {\enquote {\bibinfo {title} {A model of
  autooscillations in association reactions},}\ }\href
  {https://doi.org/10.1016/0009-2509(87)80080-5} {\bibfield  {journal}
  {\bibinfo  {journal} {Chem. Eng. Sci.}\ }\textbf {\bibinfo {volume} {42}},\
  \bibinfo {pages} {1249--1251} (\bibinfo {year} {1987})}\BibitemShut {NoStop}%
\bibitem [{\citenamefont {Stich}\ \emph {et~al.}(2013)\citenamefont {Stich},
  \citenamefont {Blanco},\ and\ \citenamefont {Hochberg}}]{dimer}%
  \BibitemOpen
  \bibfield  {author} {\bibinfo {author} {\bibfnamefont {M.}~\bibnamefont
  {Stich}}, \bibinfo {author} {\bibfnamefont {C.}~\bibnamefont {Blanco}}, \
  and\ \bibinfo {author} {\bibfnamefont {D.}~\bibnamefont {Hochberg}},\
  }\bibfield  {title} {\enquote {\bibinfo {title} {Chiral and chemical
  oscillations in a simple dimerization model},}\ }\href {\doibase
  10.1039/C2CP42620J} {\bibfield  {journal} {\bibinfo  {journal} {Phys. Chem.
  Chem. Phys.}\ }\textbf {\bibinfo {volume} {15}},\ \bibinfo {pages} {255--261}
  (\bibinfo {year} {2013})}\BibitemShut {NoStop}%
\bibitem [{\citenamefont {Matveev}\ \emph {et~al.}(2017)\citenamefont
  {Matveev}, \citenamefont {Krapivsky}, \citenamefont {Smirnov}, \citenamefont
  {Tyrtyshnikov},\ and\ \citenamefont {Brilliantov}}]{Oscil1}%
  \BibitemOpen
  \bibfield  {author} {\bibinfo {author} {\bibfnamefont {S.~A.}\ \bibnamefont
  {Matveev}}, \bibinfo {author} {\bibfnamefont {P.~L.}\ \bibnamefont
  {Krapivsky}}, \bibinfo {author} {\bibfnamefont {A.~P.}\ \bibnamefont
  {Smirnov}}, \bibinfo {author} {\bibfnamefont {E.~E.}\ \bibnamefont
  {Tyrtyshnikov}}, \ and\ \bibinfo {author} {\bibfnamefont {N.~V.}\
  \bibnamefont {Brilliantov}},\ }\bibfield  {title} {\enquote {\bibinfo {title}
  {Oscillations in aggregation-shattering processes},}\ }\href {\doibase
  10.1103/PhysRevLett.119.260601} {\bibfield  {journal} {\bibinfo  {journal}
  {Phys. Rev. Lett.}\ }\textbf {\bibinfo {volume} {119}},\ \bibinfo {pages}
  {260601} (\bibinfo {year} {2017})}\BibitemShut {NoStop}%
\bibitem [{\citenamefont {Helbing}(2001)}]{Helbing}%
  \BibitemOpen
  \bibfield  {author} {\bibinfo {author} {\bibfnamefont {D.}~\bibnamefont
  {Helbing}},\ }\bibfield  {title} {\enquote {\bibinfo {title} {Traffic and
  related self-driven many-particle systems},}\ }\href {\doibase
  10.1103/RevModPhys.73.1067} {\bibfield  {journal} {\bibinfo  {journal} {Rev.
  Mod. Phys.}\ }\textbf {\bibinfo {volume} {73}},\ \bibinfo {pages}
  {1067--1141} (\bibinfo {year} {2001})}\BibitemShut {NoStop}%
\bibitem [{\citenamefont {Friedlander}(2000)}]{Friedlander}%
  \BibitemOpen
  \bibfield  {author} {\bibinfo {author} {\bibfnamefont {S.~K.}\ \bibnamefont
  {Friedlander}},\ }\href@noop {} {\emph {\bibinfo {title} {Smoke, Dust and
  Haze}}}\ (\bibinfo  {publisher} {Oxford University Press},\ \bibinfo
  {address} {Oxford, UK},\ \bibinfo {year} {2000})\BibitemShut {NoStop}%
\bibitem [{\citenamefont {Ossenkopf}(1993)}]{Ossenkopf}%
  \BibitemOpen
  \bibfield  {author} {\bibinfo {author} {\bibfnamefont {V.}~\bibnamefont
  {Ossenkopf}},\ }\bibfield  {title} {\enquote {\bibinfo {title} {Dust
  coagulation in dense molecular clouds: The formation of fluffy aggregates},}\
  }\href {https://adsabs.harvard.edu/full/1993A%26A...280..617O} {\bibfield
  {journal} {\bibinfo  {journal} {Astron. Astrophys.}\ }\textbf {\bibinfo
  {volume} {280}},\ \bibinfo {pages} {617--646} (\bibinfo {year}
  {1993})}\BibitemShut {NoStop}%
\bibitem [{\citenamefont {Brilliantov}\ and\ \citenamefont
  {Spahn}(2006)}]{BrilliantovSpahn2006}%
  \BibitemOpen
  \bibfield  {author} {\bibinfo {author} {\bibfnamefont {N.~V.}\ \bibnamefont
  {Brilliantov}}\ and\ \bibinfo {author} {\bibfnamefont {F.}~\bibnamefont
  {Spahn}},\ }\bibfield  {title} {\enquote {\bibinfo {title} {Dust coagulation
  in equilibrium molecular gas},}\ }\href
  {https://doi.org/10.1016/j.matcom.2006.05.031} {\bibfield  {journal}
  {\bibinfo  {journal} {Math. Comput. Simul.}\ }\textbf {\bibinfo {volume}
  {72}},\ \bibinfo {pages} {93--97} (\bibinfo {year} {2006})}\BibitemShut
  {NoStop}%
\bibitem [{\citenamefont {Matveev}\ \emph {et~al.}(2015)\citenamefont
  {Matveev}, \citenamefont {Smirnov},\ and\ \citenamefont
  {Tyrtyshnikov}}]{matveev2015fast}%
  \BibitemOpen
  \bibfield  {author} {\bibinfo {author} {\bibfnamefont {S.~A.}\ \bibnamefont
  {Matveev}}, \bibinfo {author} {\bibfnamefont {A.~P.}\ \bibnamefont
  {Smirnov}}, \ and\ \bibinfo {author} {\bibfnamefont {E.~E.}\ \bibnamefont
  {Tyrtyshnikov}},\ }\bibfield  {title} {\enquote {\bibinfo {title} {A fast
  numerical method for the {C}auchy problem for the {S}moluchowski equation},}\
  }\href {https://doi.org/10.1016/j.jcp.2014.11.003} {\bibfield  {journal}
  {\bibinfo  {journal} {J. Comput. Phys.}\ }\textbf {\bibinfo {volume} {282}},\
  \bibinfo {pages} {23--32} (\bibinfo {year} {2015})}\BibitemShut {NoStop}%
\bibitem [{\citenamefont {Chaudhury}\ \emph {et~al.}(2014)\citenamefont
  {Chaudhury}, \citenamefont {Oseledets},\ and\ \citenamefont
  {Ramachandran}}]{Chaudhury2014}%
  \BibitemOpen
  \bibfield  {author} {\bibinfo {author} {\bibfnamefont {A.}~\bibnamefont
  {Chaudhury}}, \bibinfo {author} {\bibfnamefont {I.}~\bibnamefont
  {Oseledets}}, \ and\ \bibinfo {author} {\bibfnamefont {R.}~\bibnamefont
  {Ramachandran}},\ }\bibfield  {title} {\enquote {\bibinfo {title} {A
  computationally efficient technique for the solution of multi-dimensional
  pbms of granulation via tensor decomposition},}\ }\href
  {https://doi.org/10.1016/j.compchemeng.2013.10.020} {\bibfield  {journal}
  {\bibinfo  {journal} {Comput. Chem. Eng.}\ }\textbf {\bibinfo {volume}
  {61}},\ \bibinfo {pages} {234--244} (\bibinfo {year} {2014})}\BibitemShut
  {NoStop}%
\bibitem [{\citenamefont {Hackbusch}(2006)}]{hackbusch2006efficient}%
  \BibitemOpen
  \bibfield  {author} {\bibinfo {author} {\bibfnamefont {W.}~\bibnamefont
  {Hackbusch}},\ }\bibfield  {title} {\enquote {\bibinfo {title} {On the
  efficient evaluation of coalescence integrals in population balance
  models},}\ }\href {https://doi.org/10.1007/s00607-006-0174-2} {\bibfield
  {journal} {\bibinfo  {journal} {Computing}\ }\textbf {\bibinfo {volume}
  {78}},\ \bibinfo {pages} {145--159} (\bibinfo {year} {2006})}\BibitemShut
  {NoStop}%
\bibitem [{\citenamefont {Hackbusch}(2007)}]{hackbusch2007approximation}%
  \BibitemOpen
  \bibfield  {author} {\bibinfo {author} {\bibfnamefont {W.}~\bibnamefont
  {Hackbusch}},\ }\bibfield  {title} {\enquote {\bibinfo {title} {Approximation
  of coalescence integrals in population balance models with local mass
  conservation},}\ }\href {https://doi.org/10.1007/s00211-007-0077-y}
  {\bibfield  {journal} {\bibinfo  {journal} {Numer. Math.}\ }\textbf {\bibinfo
  {volume} {106}},\ \bibinfo {pages} {627--657} (\bibinfo {year}
  {2007})}\BibitemShut {NoStop}%
\bibitem [{\citenamefont {Matveev}\ \emph {et~al.}(2018)\citenamefont
  {Matveev}, \citenamefont {Ampilogova}, \citenamefont {Stadnichuk},
  \citenamefont {Tyrtyshnikov}, \citenamefont {Smirnov},\ and\ \citenamefont
  {Brilliantov}}]{matveev_CPC2018}%
  \BibitemOpen
  \bibfield  {author} {\bibinfo {author} {\bibfnamefont {S.~A.}\ \bibnamefont
  {Matveev}}, \bibinfo {author} {\bibfnamefont {N.~V.}\ \bibnamefont
  {Ampilogova}}, \bibinfo {author} {\bibfnamefont {V.~I.}\ \bibnamefont
  {Stadnichuk}}, \bibinfo {author} {\bibfnamefont {E.~E.}\ \bibnamefont
  {Tyrtyshnikov}}, \bibinfo {author} {\bibfnamefont {A.~P.}\ \bibnamefont
  {Smirnov}}, \ and\ \bibinfo {author} {\bibfnamefont {N.~V.}\ \bibnamefont
  {Brilliantov}},\ }\bibfield  {title} {\enquote {\bibinfo {title} {Anderson
  acceleration method of finding steady-state particle size distribution for a
  wide class of aggregation-fragmentation models},}\ }\href
  {https://doi.org/10.1016/j.cpc.2017.11.002} {\bibfield  {journal} {\bibinfo
  {journal} {Comput. Phys. Commun.}\ }\textbf {\bibinfo {volume} {224}},\
  \bibinfo {pages} {154--163} (\bibinfo {year} {2018})}\BibitemShut {NoStop}%
\bibitem [{\citenamefont {Strogatz}(1994)}]{Strogatz}%
  \BibitemOpen
  \bibfield  {author} {\bibinfo {author} {\bibfnamefont {S.~H.}\ \bibnamefont
  {Strogatz}},\ }\href@noop {} {\emph {\bibinfo {title} {Nonlinear Dynamics and
  Chaos}}}\ (\bibinfo  {publisher} {Addison Wesley},\ \bibinfo {address} {New
  York},\ \bibinfo {year} {1994})\BibitemShut {NoStop}%
\bibitem [{\citenamefont {French}\ \emph {et~al.}(2014)\citenamefont {French},
  \citenamefont {Hicks}, \citenamefont {Showalter}, \citenamefont {Antonsen},\
  and\ \citenamefont {Packard}}]{FRing}%
  \BibitemOpen
  \bibfield  {author} {\bibinfo {author} {\bibfnamefont {R.~S.}\ \bibnamefont
  {French}}, \bibinfo {author} {\bibfnamefont {S.~K.}\ \bibnamefont {Hicks}},
  \bibinfo {author} {\bibfnamefont {M.~R.}\ \bibnamefont {Showalter}}, \bibinfo
  {author} {\bibfnamefont {A.~K.}\ \bibnamefont {Antonsen}}, \ and\ \bibinfo
  {author} {\bibfnamefont {D.~R.}\ \bibnamefont {Packard}},\ }\bibfield
  {title} {\enquote {\bibinfo {title} {Analysis of clumps in {S}aturn's {F}
  ring from {V}oyager and {C}assini},}\ }\href
  {https://doi.org/10.1016/j.icarus.2014.06.035} {\bibfield  {journal}
  {\bibinfo  {journal} {Icarus}\ }\textbf {\bibinfo {volume} {241}},\ \bibinfo
  {pages} {200--220} (\bibinfo {year} {2014})}\BibitemShut {NoStop}%
\bibitem [{\citenamefont {Bodrova}\ \emph {et~al.}(2014)\citenamefont
  {Bodrova}, \citenamefont {Levchenko},\ and\ \citenamefont
  {Brilliantov}}]{bodrova2014}%
  \BibitemOpen
  \bibfield  {author} {\bibinfo {author} {\bibfnamefont {A.}~\bibnamefont
  {Bodrova}}, \bibinfo {author} {\bibfnamefont {D.}~\bibnamefont {Levchenko}},
  \ and\ \bibinfo {author} {\bibfnamefont {N.~V.}\ \bibnamefont
  {Brilliantov}},\ }\bibfield  {title} {\enquote {\bibinfo {title}
  {{Universality of temperature distribution in granular gas mixtures with a
  steep particle size distribution}},}\ }\href {\doibase
  10.1209/0295-5075/106/14001} {\bibfield  {journal} {\bibinfo  {journal}
  {EPL}\ }\textbf {\bibinfo {volume} {106}},\ \bibinfo {pages} {14001}
  (\bibinfo {year} {2014})}\BibitemShut {NoStop}%
\end{thebibliography}%

\end{document}